\documentclass[aps,prl,twocolumn,superscriptaddress,showpacs,floatfix]{revtex4-1}  
\usepackage{amsmath,amssymb,wasysym,graphicx}
\usepackage{times}
\usepackage[varg]{txfonts}
\usepackage{textcomp}
\usepackage{subfigure}
\usepackage{tabu}
\usepackage{color}
\usepackage{xcolor}
\usepackage[colorlinks=true,citecolor=blue,urlcolor=blue,linkcolor=blue,hyperindex]{hyperref}
\usepackage{braket}
\usepackage{dsfont}
\usepackage{overpic}
\usepackage{clrscode3e}

\usepackage[normalem]{ulem}
\usepackage{verbatim}
\allowdisplaybreaks

\begin{document}
	\title{Quantum Fisher Information as a Probe of Critical Scaling in Frustrated Magnets:\\ Signatures from Kagome Quantum Spin Liquid }

\author{Zhengbang Zhou}
\thanks{These authors contributed equally to this work.}
\affiliation{Department of Physics, University of Toronto, Toronto, Ontario M5S 1A7, Canada}

\author{Chengkang Zhou}
\thanks{These authors contributed equally to this work.}
\affiliation{Department of Physics and HK Institute of Quantum Science \& Technology, The University of Hong Kong, Pokfulam Road,  Hong Kong SAR, China}
\affiliation{State Key Laboratory of Optical Quantum Materials, The University of Hong Kong, Pokfulam Road,  Hong Kong SAR, China}

\author{Menghan Song}
\affiliation{Department of Physics and HK Institute of Quantum Science \& Technology, The University of Hong Kong, Pokfulam Road,  Hong Kong SAR, China}
\affiliation{State Key Laboratory of Optical Quantum Materials, The University of Hong Kong, Pokfulam Road,  Hong Kong SAR, China}

\author{Yong Baek Kim}
\affiliation{Department of Physics, University of Toronto, Toronto, Ontario M5S 1A7, Canada}

\author{Zi Yang Meng}
\affiliation{Department of Physics and HK Institute of Quantum Science \& Technology, The University of Hong Kong, Pokfulam Road,  Hong Kong SAR, China}
\affiliation{State Key Laboratory of Optical Quantum Materials, The University of Hong Kong, Pokfulam Road,  Hong Kong SAR, China}

\date{\today}
\begin{abstract}
Quantum Fisher information (QFI) is a measure of multipartite quantum entanglement that can be obtained from inelastic neutron scattering data on quantum magnets. In this work, we demonstrate that the QFI can distinguish an unconventional quantum critical point (QCP) with fractionalization and emergent gauge structure from conventional ones within the Landau paradigm. We compute the QFI, via large-scale quantum Monte Carlo (QMC) simulations and exact diagonalization, in a kagome lattice quantum spin liquid (QSL) model with an XY and a cluster-Ising interactions. When the XY interaction is ferromagetic, the QFI obtained by QMC reveals a large anomalous dimension, which is a fingerprint of the (2+1)d XY$^\ast$ universality class for the transition from the ferromagnetic phase to the $\mathbb{Z}_2$ QSL. The investigation of thermal and dynamical properties of QFI is further extended to the case of antiferromagnetic XY interaction via exact diagonalization. In this regime, a transition to a possibly distinct QSL phase is suggested via both entanglement-based probes, such as QFI and genuine multipartite negativity, and analyses of the energy spectrum and structure factors. These results not only demonstrate the versatility of QFI in identifying QSL states and unconventional QCPs but also provide useful guidance for future theoretical and experimental studies of frustrated magnets.


\end{abstract}
\maketitle

\noindent\textcolor{blue}{\it Introduction.}--- 
Unambiguous identification of a quantum spin liquid (QSL) in candidate materials, especially from the spectral information obtained by neutron scattering and nuclear magnetic resonance, has been a central issue in quantum magnets and quantum phase transitions therein. Representative candidate materials are now widely available for common lattice motifs such as triangular and kagome lattices~\cite{hanFractionalized2012,FuM15,FengZL17,weiEvidence2017,khuntia2020,macdougalAovided2020,ZengZ22,ghioldiEvidence2022,chatterjeeFrom2023,zengSpectral2024,scheieProximate2024,scheieSpectrum2024,bagEvidence2024,zhengUnconventional2025,hanSpin2025,breidenbackIdentifying2025}, where the continua of the excitation spectrum and quantum critical scaling behaviors at the transition from a QSL to conventional phases are expected to be used to validate the presence of a QSL~\cite{hanFractionalized2012,FuM15,FengZL17,weiEvidence2017,khuntia2020,chatterjeeFrom2023,scheieProximate2024,zengSpectral2024,bagEvidence2024,breidenbackIdentifying2025}. 
However, quantitative comparisons between model computations and experimental results remain rare, making it challenging to understand the precise connection between the nature of fractionalization and emergent gauge structures of QSL phases in theoretical~\cite{SavaryL17,ZhouY17,wenChoreographed2019,broholm2020Quantum} model computations~\cite{yanSpin2011,DepenbrockS12,liaoGapless2017,Dynamical2018Sun,Dynamics2018Huang} and those hidden in real experimental data in candidate materials.

The difficulty in making a quantitative comparison arises from both theoretical and experimental challenges. On the one hand, the currently available analytical and numerical methods do not yet provide a quantitatively reliable description of their dynamical responses across the experimentally relevant regimes (e.g., low temperatures, applied fields, and momentum–frequency resolution), particularly for triangular and kagome antiferromagnets. On the other hand, the excitation spectra obtained are often described by a continuum of two-spinon excitations, but the experimental probes may not have enough energy and momentum resolution to identify smoking gun signatures in the data and distinguish them from other possible mechanisms, such as disorder.



\begin{figure}[htp!]
\centering
\includegraphics[width=\columnwidth]{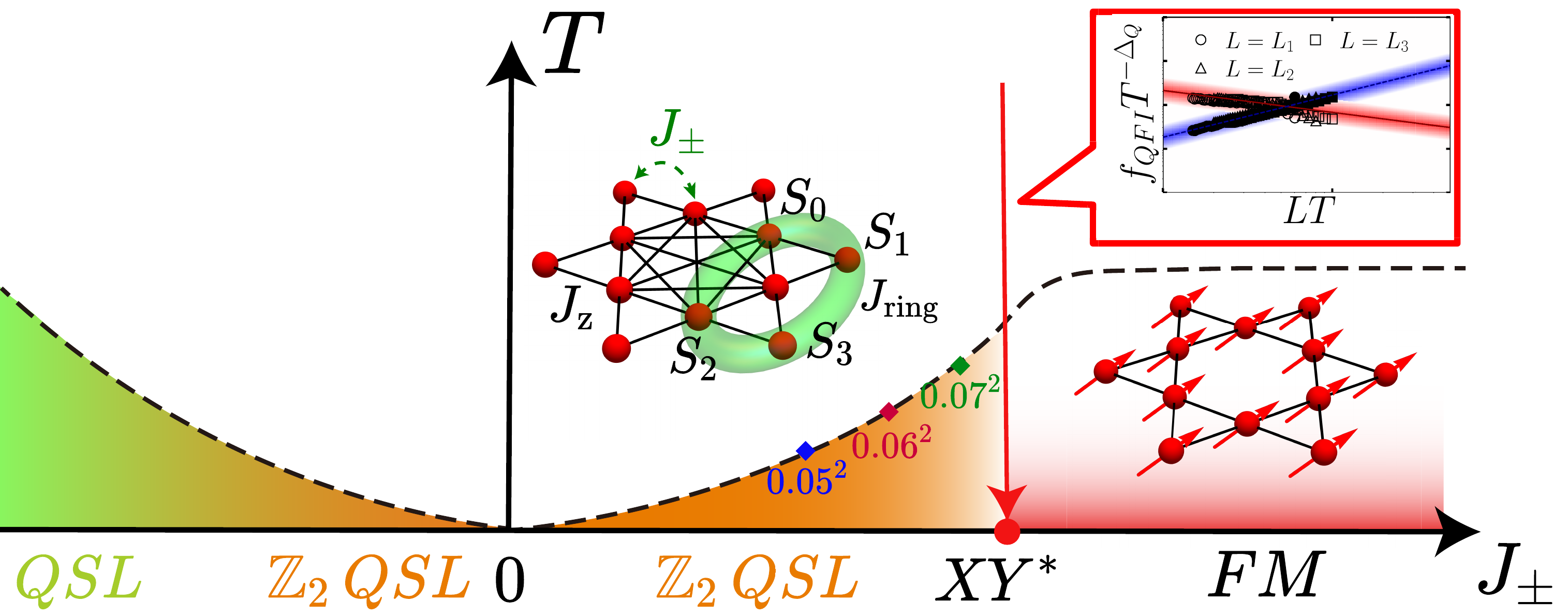}
\caption{\textbf{Model, phase diagram and the dynamical scaling of QFI in kagome QSL.} 
The schematic phase diagram of the kagome QSL model shows the $\mathbb{Z}2$ QSL phase (orange region) and the ferromagnetic (FM) phase (red region). The color gradient within the red region indicates that the FM phase only appears at zero temperature. The nearest-neighbor hopping term $J{\pm}$ and the coupling $J_z=1$ (taken as the energy unit) between each pair of spins within the same hexagon are depicted. The black dashed line traces the location of the lowest-temperature peak in the specific heat based on ED results in SM~\cite{suppl}. The green circle in the left inset refers to the effective four-spin exchange coupling $J_{\mathrm{ring}}$ in the $\mathbb{Z}2$ QSL phase, which sets the temperature scale for entering the QSL phase $T \sim J_{\mathrm{ring}} \sim J_{\pm}^2$~\cite{Fractionalization2002Balents}, as examplified by the blue, red, and green diamonds for $J_{\pm}=0.05$, $0.06$, and $0.07$, respectively.
The critical point (red dot) between the FM and $\mathbb{Z}2$ QSL phases belongs to the (2+1)d XY$^\ast$ universality class. The upper right inset illustrates how the quantum Fisher information (QFI) scaling behaves at this critical point: as temperature decreases, it captures the anomalous scaling dimension characteristic of the XY$^\ast$ transition (red shaded data) rather than that of the conventional XY transition (blue shaded data).
Additionally, our numerical results in the $J{\pm}<0$ regime identify a possible QSL region (green region) distinct from the $\mathbb{Z}_2$ QSL (orange region).}
	\label{fig:lattice}
\end{figure}

In this letter, we demonstrate that the quantum Fisher information (QFI)~\cite{lambert2019estimates, lambert2020revealing, Witnessing2021Scheie, Proximate2024scheie, laurell2021quantifying, menon2023multipartite, Tutorial2025Scheie, laurell2025witnessing, Can2025shimokawa,hyllus2012fisher, toth2012multipartite, Witnessing2021Scheie, Tutorial2025Scheie, liu2016quantum, zhouQuantum2025,Sabharwal2025Characterizing,Entanglement2009Smerzi}, a measure of multipartite quantum entanglement~\cite{Entanglement2025Song,Multiparty2025Lyu,Long-range2025Avakian,Detecting2005Doherty,wangEntanglement2025,wangAnalog2025}, may help overcome some of the difficulties mentioned above.
The QFI provides a lower bound on the multipartite entanglement in the system, and is directly related to the dynamical spin structure factor~\cite{hauke2016measuring}, which is routinely measured in inelastic neutron scattering experiments~\cite{lovesey1984theory, boothroyd2020principles, bramwell2014neutron}. For example, the scaling behavior of the QFI has been argued to probe multipartite entanglement in 1d systems~\cite{hauke2016measuring,Witnessing2025Sabharwal,Witnessing2021Scheie}. More recently, the temperature and momentum dependence of the QFI has been used to elucidate the thermal and dynamical properties of a pyrochlore lattice model of quantum spin ice, a 3d QSL hosting fractionalized quasiparticles and emergent photons~\cite{zhouQuantum2025}. Furthermore, it is demonstrated that QFI can be viewed as an experimentally accessible measure to distinguish a QSL from a competing disorder-driven random singlet phase on the triangular lattice~\cite{Can2025shimokawa}.

\begin{figure*}[htp!]
	\centering
    \includegraphics[width=\textwidth]{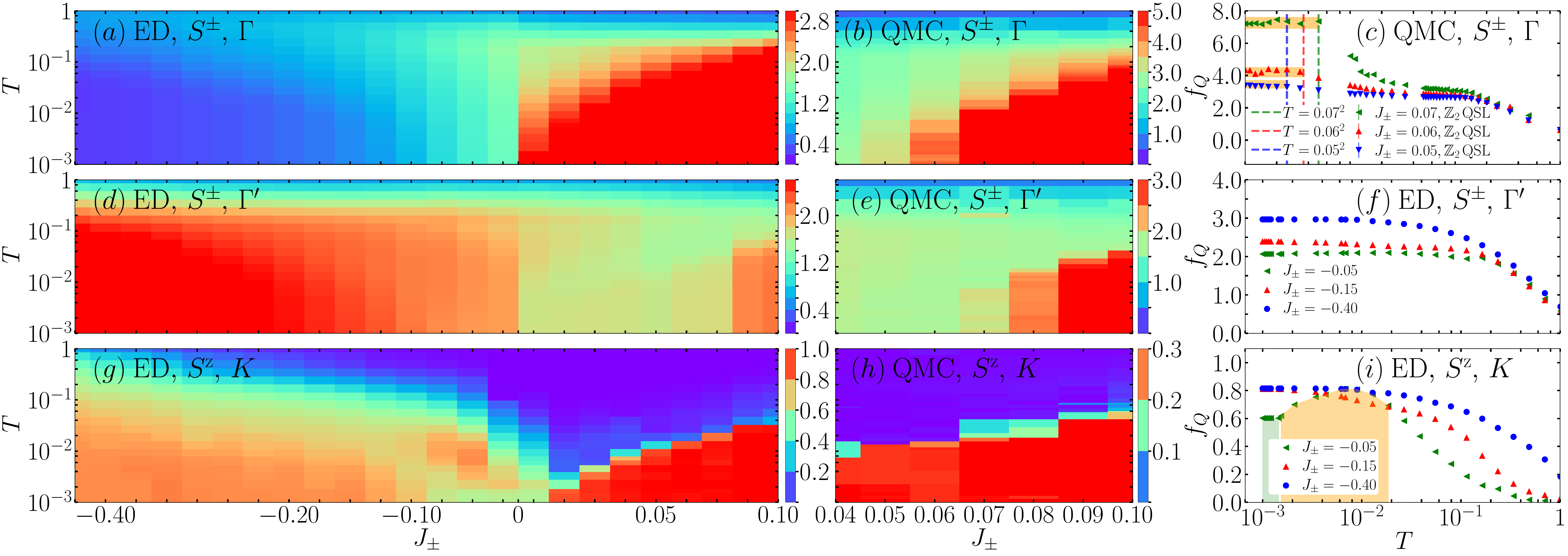}
	\caption{\textbf{QFI map in the Kagome spin liquid model.} The QFI density $f_Q$ obtained from ED and QMC is shown as a function of $J_{\pm}$ and temperature $T$. The system size is $N=3\times L_x \times L_y=3\times 3 \times 3=27$ for ED (panels (a), (d) and (g)) and $N=3\times 6\times 6$ for QMC (panels (b), (e) and (h)). Panels (a) and (b) show $f_Q$ in the $S^{\pm}$ channel at $\mathbf{q}=\Gamma=(0,0)$, mapping out the thermal and quantum phase diagram of the BFG model and revealing the phase boundary between the FM and $\mathbb{Z}_2$ QSL phases. Panels (d) and (e) display $f_Q$ at $\mathbf{q}=\Gamma^{\prime}=(2\pi,2/\sqrt{3}\pi)$ in the same channel, reflecting the strength of fluctuations across the thermal and quantum phase diagram. Panel (c) shows line cuts of (b) for $J_{\pm}=0.05$, $0.06$, and $0.07$, where the dashed line represents the temperature scale for entering the QSL phase, determined by $T \sim J_{\mathrm{ring}} \sim J_{\pm}^2$. The $\mathbb{Z}_2$ QSL phase is highlighted in orange shading. Panel (f) shows line cuts of (e) for $J_{\pm}=-0.05$, $-0.15$, and $-0.40$ for the detailed QFI behavior with decreasing temperature in the AFM regime. Furthermore, panels (g) and (h) show $f_Q$ in the $S^z$ channel at $\mathbf{q}=K=(4\pi/3,0)$ for ED and QMC, respectively. Panel (g) reveals a region with a characteristic hump around $T\approx10^{-2}$ for $J_{\pm}\in[-0.15,0]$ and reaches a lower plateau at lower temperature. Panel (i) shows line cuts of (g) for $J_{\pm}=-0.05$, $-0.15$, and $-0.40$, with the hump structure at $J_{\pm}=-0.05$ highlighted in orange shading and the lower plateau in green shading. 
	}
	\label{fig:qfi_Full}
\end{figure*}

In this work, we explore the QFI as a tool to identify the QSL state in 2d and its non-Landau QCP arising from fractionalization and emergent gauge field for the transition to a conventional phase. Here we show, by means of large-scale quantum Monte Carlo (QMC) and exact diagonalization (ED) simulations in a kagome lattice QSL model with an XY and a cluster-Ising interactions, that QFI can indeed identify such a novel QCP. Our results on thermal scaling behavior of the QFI in the case of ferromagnetic XY interaction, clearly capture the anomalous scaling dimension of the (2+1)d XY$^\ast$ transition from a ferromagnetic phase to $\mathbb{Z}_2$ QSL in contrast to the (2+1)d XY one that falls into the Landau paradigm. 

Moreover, the investigation of thermal and dynamical properties of QFI is further extended to the antiferromagnetic XY interaction regime, which may be more relevant to the experimental progress in finding QSL in candidate materials~\cite{hanFractionalized2012,FuM15,FengZL17,weiEvidence2017,khuntia2020,heringPhase2022,chatterjeeFrom2023,zengSpectral2024,hanSpin2025,liAntiferromagnetic2025,breidenbackIdentifying2025}. In this regime, we also consider the genuine multipartite negativity (GMN)~\cite{Guhne2011Taming,Hofmann2014Analytical,Entanglement2025Song,lyu2025multiparty} calculation to probe the multipartite entanglement structure. Specifically, the minimal multipartite entangled subregion describes the smallest subregion that hosts the non-zero GMN signal in the system, and the change of this subregion indicates the change of the entanglement structure in the system. Using the QFI and GMN, we find evidence of a transition from the $\mathbb{Z}_2$ QSL to a novel QSL in the antiferromagnetic XY interaction regime, which has not been identified in earlier studies. The possible presence of this transition is further supported by the energy spectrum in the ED calculation, which shows a change in relevant momentum sectors from $\mathbb{Z}_2$ QSL to the other phase.

\noindent\textcolor{blue}{\it Model and Method.}--- We consider the Balents-Fisher-Girvin (BFG) model \cite{Fractionalization2002Balents,isakovSpinLiquid2006,isakovTopological2011,isakovUniversal2012,wangTopological2017,Dynamical2018Sun,QSL2018Wang,Fractionalized2021wang} on the kagome lattice, whose Hamiltonian is given by
\begin{equation}
H = -J_{\pm}\sum_{\langle i,j \rangle} (S_i^+ S_j^- + S_i^- S_j^+) + \frac{J_z}{2} \sum_{\hexagon} \left( \sum_{i \in \hexagon} S_i^z \right)^2.
\label{eq:Hamiltonian}
\end{equation}
As illustrated in Fig.~\ref{fig:lattice}, the first term describes the nearest-neighbor XY interaction with coupling strength $J_{\pm}$, while the second term is the cluster interaction between pairs of spins in each hexagon $\hexagon$ of the kagome lattice with coupling strength $J_z$. In our simulations, we set $J_z=1$ as the energy unit. The phase diagram for $J_{\pm}>0$ has been numerically investigated in QMC~\cite{isakovSpinLiquid2006,isakovTopological2011,isakovUniversal2012,wangTopological2017,Dynamical2018Sun,QSL2018Wang}. There is a $\mathbb{Z}_2$ QSL phase in the regime of $J_{\pm}<0.07076$ and a ferromagnetic (FM) phase for $J_{\pm}>0.07076$. 
 In the $\mathbb{Z}_2$ QSL phase, the effective four-spin exchange coupling $H_{\mathrm{eff}}=-J_{\mathrm{ring}}(S_0^+S_1^-S_2^+S_3^-+\mathrm{h.c.})$ is indicated by green circles in Fig.\ref{fig:lattice}, where $J_{\mathrm{ring}}\sim J_{\pm}^2$. The temperature scale characterizing the QSL phase is set by $T \sim J_{\mathrm{ring}} \sim J_{\pm}^2$, with examples $T \sim (0.05)^2$ (blue diamond), $(0.06)^2$ (red diamond), and $(0.07)^2$ (green diamond) shown in Fig.\ref{fig:lattice}, respectively.
 At zero temperature, the phase transition between $\mathbb{Z}_2$ QSL and FM phase is continuous and belongs to the (2+1)d XY$^\ast$ universality class \cite{isakovUniversal2012,Dynamical2018Sun,QSL2018Wang}, with the anomalous scaling dimension of the $S^{\pm}$ operator $\Delta=\frac{1+\eta}{2}=\frac{\beta}{\nu}\sim1.25$~\cite{Fractionalization2002Balents,isakovUniversal2012,QSL2018Wang,Fractionalized2021wang} instead of that of the conventional (2+1)d XY transition $\Delta=\frac{1+\eta}{2}=\frac{\beta}{\nu}\sim0.52$~\cite{mengPhase2008,hohenadlerQuantum2012}, precisely due to the fractionalization of $S=1$ magnon to $S=1/2$ spinon and the associated visons and the emergent $\mathbb{Z}_2$ gauge field at the QCP~\cite{Fractionalization2002Balents,isakovUniversal2012,QSL2018Wang,Fractionalized2021wang}. 
In contrast, for $J_{\pm}<0$, the phase diagram remains to be explored due to the sign-problem in QMC. 

To probe the full phase diagram, we employ ED for both $J_{\pm}<0$ and $J_{\pm}>0$, and QMC for $J_{\pm}>0$. The ED simulations use clusters of sizes $L_x=L_y=3$ with total $N=3\times L_x\times L_y=27$ spins, while QMC simulations access larger systems with $L_x=L_y=6$ to $12$. Both methods employ periodic boundary conditions and calculate the QFI density $f_Q$ at different momentum points $\mathbf{q}$, defined as
\begin{equation}
	f_{Q}(S^\alpha_\mathbf{q}, T)=4 \int_{0}^{\infty} d \omega \tanh \left(\frac{\omega}{2T}\right)\left(1-e^{- \omega/T}\right) A^\alpha(\mathbf{q}, \omega),
	\label{eq:QFI}
\end{equation}
where $A^\alpha(\mathbf{q}, \omega)$ is the dynamical spin structure factor (DSSF) defined as $A^\alpha(\mathbf{q}, \omega):=\frac{1}{2\pi N}\int dt\langle S^{\alpha\dagger}_\mathbf{q}(t)S^\alpha_{\mathbf{q}}(0)\rangle e^{i\omega t}$ at momentum $\mathbf{q}$. Here, $T$ is the temperature and $\alpha \in\{\mathrm{x},\mathrm{y},\mathrm{z}\}$ labels different components. To obtain the DSSF from QMC simulations, we measure the imaginary-time spin-spin correlation functions and then perform stochastic analytic continuation (SAC) \cite{Stochastic1998sandvik,Using2008syliuaasen,Identifying2004beach,Constrained2016sandvik,Nearly2017shao,Amplitude2021zhou,zhouQuantum2025} to extract the DSSF from the imaginary-time data. For the ED simulations, besides the typical ground state results, the scheme of $f_Q$ follows the microcanonical thermal pure quantum method (mTPQ)~\cite{thermal2012sugiura, canonical2013sugiura}, and the DSSF is calculated via Lanczos approximation of the continued fraction form of the Green's function~\cite{gagliano1987,hallberg1995,prelovsek2013lanczoschapter}. Technical details of both methods are provided in Secs. I and II of the Supplemental Material (SM)~\cite{suppl}.

Moreover, to further probe the multipartite entanglement structure in the AFM regime, we compute the GMN~\cite{Guhne2011Taming,Hofmann2014Analytical,Entanglement2025Song,lyu2025multiparty}, an entanglement monotone that quantifies genuine multipartite entanglement. The GMN builds on the bipartite negativity: for any bipartition $M|\overline{M}$, the negativity is defined as ${N}_{M|\overline{M}}(\rho) = \frac{1}{2} ( |\rho^{T_M}|_1 - 1 )$, where $T_M$ denotes partial transposition with respect to $M$ and $|\cdot|_1$ is the trace norm. 
GMN probes multipartite entanglement in specific real-space subregions and can exhibit singular behavior near quantum phase transitions~\cite{wangEntanglement2025,Multiparty2025Lyu}. Recent work has further shown that GMN can vanish on non-loopy subregions in certain collectively entangled phases, including QSLs~\cite{lyu2025multiparty,Sabharwal2025Characterizing} and beyond-Landau quantum critical points~\cite{Entanglement2025Song}.
Specifically, we evaluate GMN as a function of $J_{\pm}$ on four representative subregions defined in the bowtie and hexagon. Detailed definitions and computational procedures are provided in Sec. III of the SM~\cite{suppl}.


\medskip

\noindent\textcolor{blue}{\it Results.}---Our results for the QFI density $f_Q$ in the channel $f_Q(S^\pm_{\mathbf{q}},T) = f_Q(S^\mathrm{x}_{\mathbf{q}},T) + f_Q(S^\mathrm{y}_{\mathbf{q}},T)$ and $f_Q(S^\mathrm{z}_{\mathbf{q}},T)$ are plotted in Fig.~\ref{fig:qfi_Full} as a function of both $J_{\pm}$ and temperature $T$ at the momenta $\Gamma=(0,0)$ ([Panels (a-c)]) and $\Gamma^\prime=(2\pi,2/\sqrt{3}\pi)$ (Panels (d-f)) for the $S^\pm$ channel, and at $K=(4\pi/3,0)$ for the $S^z$ channel (Panels (g-i)). Panels (b), (c), (e), and (h) are obtained from QMC for $J_{\pm} > 0$, while the others are from ED calculations. 

At the $\Gamma$ point, as shown in panels (a) and (b), the QFI density $f_Q$ serves as an effective probe to distinguish different thermal phases in the BFG model. In panel (c), we plot the temperature dependence of the QFI within the $\mathbb{Z}_2$ QSL regime for $J_{\pm} = 0.05, 0.06, 0.07$. The QFI density $f_Q$ is small at high temperatures but increases to a plateau as the temperature decreases. As the temperature decreases further, $f_Q$ increases again and reaches a higher plateau (highlighted by the yellow shading) below the energy scale of the effective four-spin ring-exchange coupling $T \sim J_{\mathrm{ring}} \sim J_{\pm}^2$ (marked by the dashed line in panel (c)). This behavior indicates the emergence of strong multipartite entanglement when entering the $\mathbb{Z}_2$ QSL phase. On the other hand, in the FM regime ($J_{\pm}/J_z>0.07076$), the QFI density $f_Q$ also increases as the temperature decreases. It comes from the onset of the enhanced FM order at low temperatures in finite size systems. We note, in the thermodynamic limit, the true FM order will only occur at zero temperature.

Meanwhile, in the $J_{\pm}<0$ regime, the QFI density $f_Q$ extracted from ED reveals a thermal phase diagram qualitatively distinct from that for $J_{\pm}>0$. As shown in Fig.~\ref{fig:qfi_Full}(d--f), $f_Q(S^\pm_{\Gamma'},T)$ increases upon cooling within the QSL regime. However, unlike the $J_{\pm}>0$ case, no well-defined intermediate plateau appears for stronger AFM couplings, such as $J_{\pm}=-0.05$ and $-0.15$ [green and red points in Fig.~\ref{fig:qfi_Full}(f)]. A sharper distinction emerges in the $S^z$ channel at the $K$ point [Fig.~\ref{fig:qfi_Full}(g--i)]: for weak AFM coupling ($-0.1<J_{\pm}<0$), $f_Q$ develops a pronounced hump near $T\approx10^{-2}$ [orange shading in Fig.~\ref{fig:qfi_Full}(i) for $J_{\pm}=-0.05$], followed by a crossover to a lower-temperature plateau [green shading in Fig.~\ref{fig:qfi_Full}(i)]. Such a hump structure is not observed in the $J_{\pm}>0$ regime and vanishes as $J_{\pm}$ decreases further, e.g., $J_{\pm}=-0.15$ and $-0.4$, as shown in Fig.~\ref{fig:qfi_Full}(i). Meanwhile, the lower plateau value of $f_Q$ also increases as $J_{\pm}$ decreases, and finally reaches a higher value ($\approx0.81$) for strong AFM $J_{\pm}$, e.g., $J_{\pm}=-0.4$. Taken together, these trends may suggest the onset of a distinct phase, different from the $\mathbb{Z}_2$ QSL close to $J_{\pm}=0$ limit, near $J_\pm \sim -0.1$.

To further investigate the phase diagram, we first focus on the scaling behavior of the QFI at the critical point in the $J_\pm>0$ regime. After which, we leverage various diagnostics to probe the nature of the possible distinct phase in the $J_\pm<0$ regime.

\noindent\textcolor{blue}{\emph{(i) The $J_\pm >0$ regime}} ---Here, we investigate the scaling behavior of the QFI density $f_Q$ at the critical point $J_{\pm,c}=0.07076$ as a function of temperature ($T$) with different system sizes $L$, which reflects the anomalous scaling dimension of the (2+1)d XY$^\ast$ transition. Previous works~\cite{hauke2016measuring,menon2023multipartite,Witnessing2021Scheie} proposed that at the critical point, the QFI density scales as
\begin{equation}
f_Q=\lambda^{\Delta_Q}\phi(T\lambda^z,L^{-1} \lambda,h\lambda^{1/\nu}),
\label{eq:eq3}
\end{equation}
where $\phi$ is a universal function of its dimensionless arguments, and $\lambda$ is the finite-size space-time cutoff scale at the critical point. $\nu$ and $z$ are the correlation length and dynamical critical exponents, respectively. $L$ is the linear system size, and $h$ is the strength of a conjugate field that drives the system away from criticality. In our case, $h=\frac{J_{\pm}}{J_z}-(\frac{J_\pm}{J_z})_c$.

The scaling exponent of the QFI is $\Delta_Q = d - 2\Delta$, where $\Delta$ is the scaling dimension of the order parameter operator ($S^\pm$) and $d$ is the spatial dimension. At the critical point of (2+1)d XY$^\ast$ universality class, i.e. $h=0$, $z=1$ and $\Delta_Q = d - 2\Delta=2-(1+\eta)=-0.495$. In sharp contrast, if the phase transition belongs to the (2+1)d XY class, i.e. $h=0$, $z=1$ and $\Delta_Q = d - 2\Delta=2-(1+\eta)=0.96$. Therefore, by setting $\lambda=L$ at low temperature ($L\ll 1/T$) and fixing the temperature as $T=1/(bL)$ with a constant $b$, the scaling form of the QFI density at the critical point becomes $f_Q = L^{\Delta_Q} \phi(1/b)$. It means that if we plot $f_Q$ as a function of $L$ at the critical point with fixed $T=1/(bL)$, all the data should follow a power-law behavior with power $\Delta_Q=-0.495$ if the transition is of beyond-Landau type, and with power $\Delta_Q=0.96$ if the transition is of Landau type. Alternatively, we can obtain the finite-size scaling form of the QFI density at the critical point as $f_Q T^{\Delta_Q} = (TL)^{\Delta_Q} \phi(TL)$. It means that if we plot $f_Q T^{\Delta_Q}$ as a function of $TL$ for different system sizes $L$ at the critical point, all the data should collapse onto a single curve. The slope of the universal curve, will distinguish the beyond-Landau transition (with negative slope) from the Landau-type transition (with positive slope).

\begin{figure}
	\centering
	\includegraphics[width=\columnwidth]{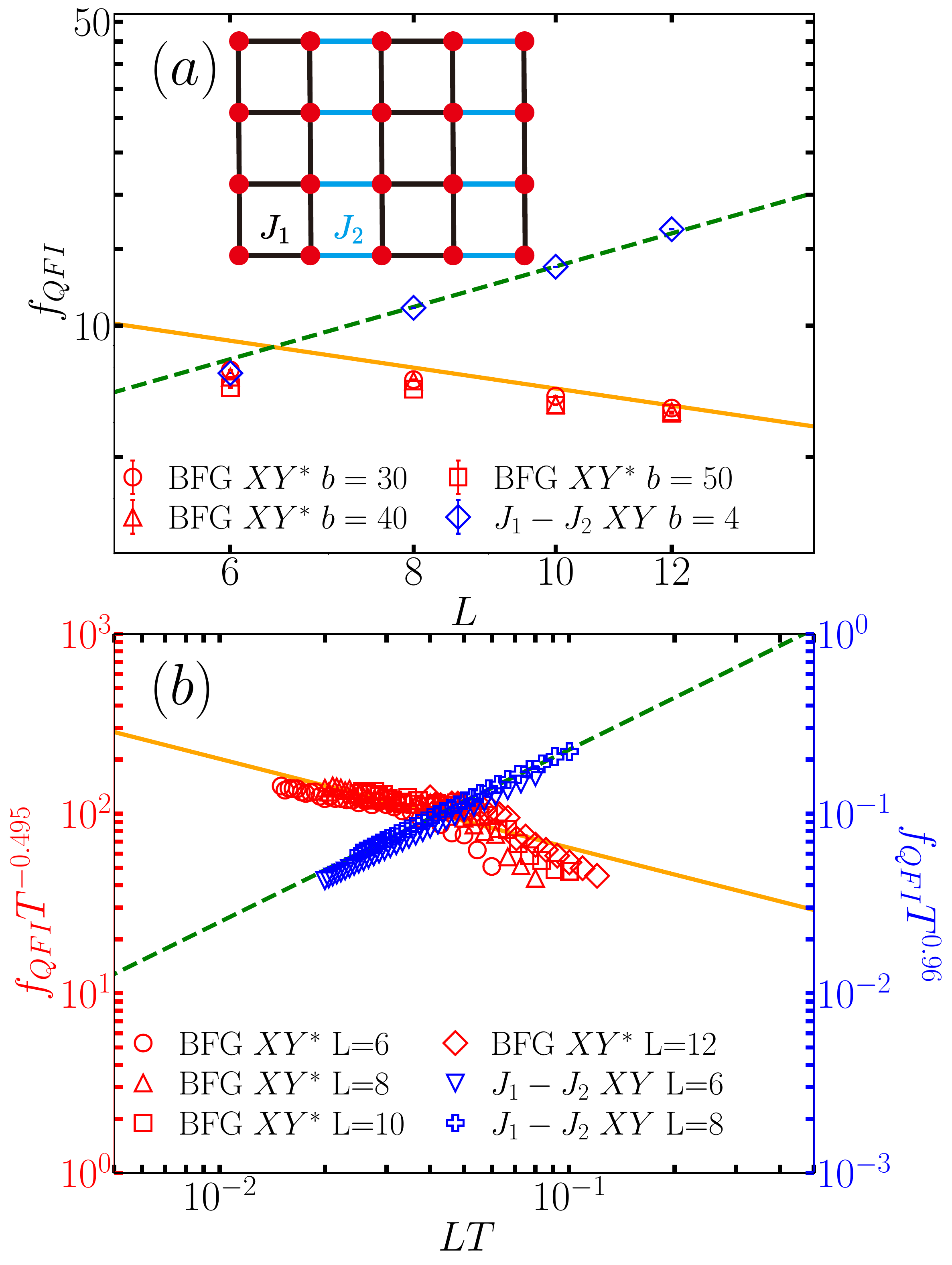}
	\caption{\textbf{Dynamical scaling of QFI to identify beyond-Landau QCP with fractionalization and emergent gauge field.} (a) The scaling behavior of QFI density $f_Q(\Gamma)$ is plotted as a function of temperature $T$ with fixed $T=1/(bL)$ at the critical point ($h=0$) for different system sizes ranging from $L=6$ to $L=12$, where $b$ ranges from $30$ to $50$. And the inset figure is the schematic figure of the easy-plane $J_1$-$J_2$ model. (b) shows the scaling behavior of $f_Q(\Gamma) T^{\Delta_Q}$ as a function of $LT$ at the critical point for different system sizes ranging from $L=6$ to $L=12$ with $\beta\in[100,400]$. In both panels, the red points refer to the BFG model, whose phase transition belongs to the (2+1)d XY$^\ast$ universality class -- beyond-Landau QCP, while the blue points refer to the easy plane $J_1$-$J_2$ model, whose phase transition belongs to the (2+1)d XY universality class. In panel (b), the red (left) y-axis refers to the XY$^{\ast}$ case while the blue (right) one is for the XY case. The orange solid lines correspond to a power-law with exponent $1+\eta=-0.495$ (XY$^{\ast}$ QCP), while the green dashed lines correspond to an exponent of $0.96$ (XY QCP). Both panels clearly show the different scaling behaviors of QFI density $f_Q$ at the critical point between the (2+1)d XY$^\ast$ universality class and the (2+1)d XY universality class.
	}
	\label{fig:scaling}
\end{figure}


To compare these two universality classes, we also perform QMC simulations on the easy-plane $J_1$-$J_2$ model on the square lattice~\cite{mengPhase2008,hohenadlerQuantum2012,Dynamical2018ma,Entanglement2025Song}, whose Hamiltonian is given by
\begin{equation}
\begin{aligned}
H_{J_1J_2} =& J_1 \sum_{\langle i,j \rangle} (S_i^x S_j^x + S_i^y S_j^y + \Delta S_i^z S_j^z) \\&+ J_2 \sum_{\langle\langle i,j \rangle\rangle} (S_i^x S_j^x + S_i^y S_j^y + \Delta S_i^z S_j^z),
\end{aligned}
\label{eq:J1J2}
\end{equation}	
As shown in the inset of Fig.~\ref{fig:scaling}(a), the black bonds denote the $J_1$ coupling and the blue bonds denote the $J_2$ coupling. We use XXZ-type couplings by setting $\Delta=1/2$. By tuning the ratio $J_1/J_2$, a phase transition from the AFM phase to the paramagnetic phase occurs at $J_1/J_2=2.735(2)$ and belongs to the (2+1)d XY universality class~\cite{Dynamical2018ma,Entanglement2025Song}. Detailed simulation results are provided in the SM~\cite{suppl}.

Our data are presented in Fig.~\ref{fig:scaling}. Panel (a) shows the scaling behavior of $f_Q$ as a function of system size $L$ with fixed $T=1/(bL)$ at the critical point for the BFG model (red points) at $J_{\pm,c}=0.07076$ ($h=0$) with $b$ ranging from $30$ to $50$. For comparison, the scaling behavior of QFI for the easy-plane $J_1$-$J_2$ model is plotted as the blue points with $J_1/J_2=2.735(2)$ and $b=4$. As $L$ increases, both data sets follow a power-law behavior. However, the BFG model data (red points) approach the power-law behavior with exponent $-0.495$ (orange solid line), while the easy-plane $J_1$-$J_2$ model data (blue points) approach the power-law behavior with exponent $0.96$ (green dashed line). Such a clear distinction in the scaling behavior of $f_Q$ at the critical point between the BFG model and the easy-plane $J_1$-$J_2$ model is caused by 
the emergence of the fractionalized spinon excitations and the associated $\mathbb{Z}_2$ gauge field at the (2+1)d XY$^\ast$ QCP in the BFG model, which leads to the anomalous scaling dimension of the $S^{\pm}$ operator.

Furthermore, Fig.~\ref{fig:scaling}(b) shows the second scaling behavior of $f_Q T^{\Delta_Q}$ as a function of $LT$ at the critical point for different system sizes: from $L=6$ to $L=12$ with $\beta\in[100,400]$ for the BFG model (red points), and from $L=8$ to $L=10$ for the easy-plane $J_1$-$J_2$ model (blue points). As $LT$ decreases, the $f_Q$ values for these system sizes collapse onto a single curve, following a power law with an exponent of $-0.495$ (orange solid line), which decreases as a function of $LT$. In sharp contrast, the QFI in the easy-plane $J_1$-$J_2$ model follows a power law with an exponent of $0.96$ (green dashed line), which increases as a function of $LT$. Once again, the opposite scaling behaviors of $f_Q$ at the critical point between the BFG model and the easy-plane $J_1$-$J_2$ model further confirm that their phase transitions belong to different universality classes.

The uniqueness of the XY$^\ast$ transition is manifested in its anomalous scaling dimension of the $S^{\pm}$ operator, and our results confirm that the QFI density $f_Q$ at the critical point indeed captures the anomalous scaling dimension of the (2+1)d XY$^\ast$ transition instead of that of the conventional (2+1)d XY one, demonstrating that the QFI can identify the novel QCP with fractionalization and emergent gauge structure. We note that the enhancement of the scaling dimension caused by the fractionalization is a universal phenomenon and has also been recently observed in quantum loop models~\cite{ranHidden2024,ranCubic2024} with implications for the persistence of fractionalization at finite temperatures~\cite{ranPhase2025}. 

\begin{figure}
	\centering
	\includegraphics[width=\columnwidth]{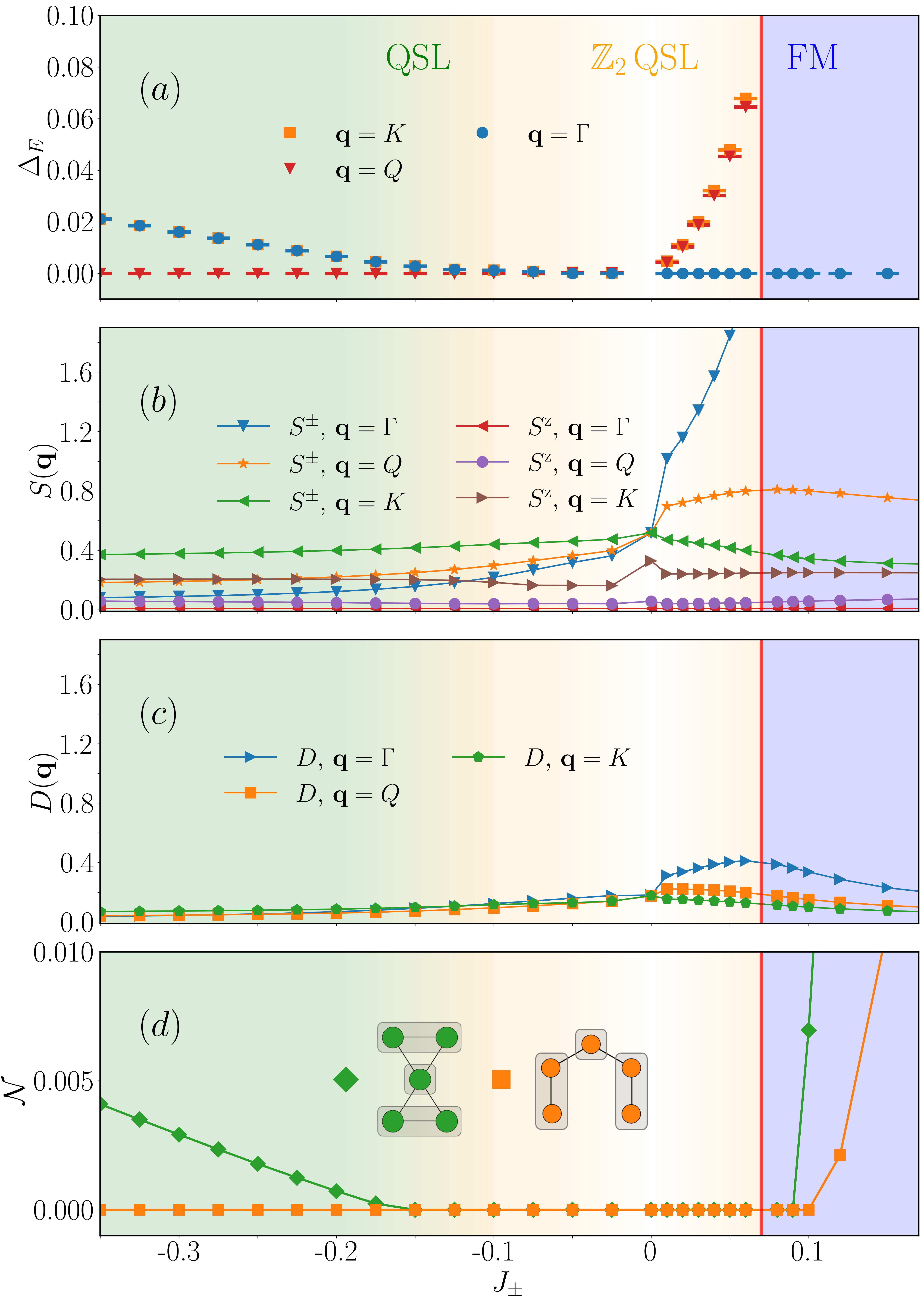}
  	\caption{\textbf{ED results on the AFM kagome model.}
	(a) shows the energy gap $\Delta E = E_q - E_0$ as a function of $J_{\pm}$ at zero temperature for the same system size, where $E_0$ is the ground-state energy and $E_q$ is the lowest energy level in the momentum sector $q$, with $q$ taking the values $\Gamma=(0,0)$, $K=(4\pi/3,0)$, $Q=(0,4\sqrt{3}\pi/9)$, and other symmetry-related points. (b) shows the static spin structure factor $S(\mathbf{q})$ at these momenta as a function of $J_{\pm}$, while (c) shows the dimer structure factor $D(\mathbf{q})=\frac{1}{N}\sum_{i,j} e^{i\mathbf{q}(\mathbf{r}_i-\mathbf{r}_j)}\langle D_iD_j\rangle- \langle D_i \rangle\langle D_j\rangle $ where $D_i=\mathbf{S}_i\cdot \mathbf{S}_{i+\hat{x}}$ for the nearest-neighbor spins at these same momenta as a function of $J_{\pm}$; both (b) and (c) are at the zero temperature for $3\times 3$ cluster as in (a). (d) shows the GMN as a function of $J_{\pm}$ for the same system size and for different partitions obtained from ED calculations. The inset of (d) shows a zoomed view near $J_{\pm}=-0.3$. In all panels, three different regimes are highlighted with different colors: green shading for the possible QSL phase, blue shading for the FM phase, and orange shading for the $\mathbb{Z}_2$ QSL phase, with an inner white region indicating the vicinity of the classical limit at $J{\pm}=0$. The red solid line marks the critical point $J_{\pm,c}=0.07076$ between the FM and $\mathbb{Z}_2$ QSL phases in the $J{\pm}>0$ regime.
	}
	\label{fig:fig4}
\end{figure}

\medskip
\noindent\textcolor{blue}{\emph{(ii) The $J_\pm<0$ regime.}}---To investigate the potentially distinct phase indicated in the QFI map [Fig.~\ref{fig:qfi_Full} (f),(g),(h), (i)], we compute the energy spectrum, static spin structure factor $S(\mathbf{q})$, and dimer structure factor $D(\mathbf{q})$ from ED calculations to examine possible energy crossings and conventional order parameters. In addition, the GMN~\cite{Guhne2011Taming, Hofmann2014Analytical, Entanglement2025Song, lyu2025multiparty} is utilized to further characterize the entanglement structure. The results are shown in Fig.~\ref{fig:fig4}.

Figure~\ref{fig:fig4} (a) shows the energy gap $\Delta E = E_q - E_0$ as a function of $J_{\pm}$ at zero temperature for $3\times3$ cluster, where $E_0$ is the ground-state energy and $E_q$ is the lowest energy level in the momentum sector $q$, with $q$ taking the values $\Gamma=(0,0)$, $K=(4\pi/3,0)$, $Q=(0,4\sqrt{3}\pi/9)$ together with other symmetry-related momenta. The energy spectrum reveals three distinct regimes. For $J_{\pm}>0.07076$, the system enters the FM phase (blue shaded region), while it is in the $\mathbb{Z}_2$ QSL phase (orange shaded region) for $-0.1 <J_{\pm}<0.07076$. These two regimes are separated by a critical point at $J_{\pm,c}=0.07076$ (red solid line). For $J_{\pm}<0$, the energy spectrum exhibits a different behavior: the lowest energy levels in the $\Gamma$, $K$, and $Q$ sectors remain nearly degenerate until $J_{\pm} \sim -0.1$, where the $\Gamma$ and $K$ sectors begin to split from the $Q$ sector. In this region, the system remains in the $\mathbb{Z}2$ QSL phase~\cite{wangExtended2009}. As $J{\pm}$ decreases further, the $Q$ sector becomes the lowest energy level, while the $\Gamma$ and $K$ sectors remain nearly degenerate. This change in the low-energy spectrum suggests a possible rearrangement of the ground state structure, consistent with the QFI map, which indicates a distinct phase for $J_{\pm}<-0.1$ (green shaded region).

To further investigate whether any conventional order exists in this distinct phase, we also measure the spin structure factor $S(\mathbf{q})$ in both the $S^{\pm}$ and $S^{z}$ channels [see Fig.~\ref{fig:fig4}(b)], as well as the dimer structure factor $D(\mathbf{q})$ [see Fig.~\ref{fig:fig4}(c)] at the relevant momentum points (see details in the SM~\cite{suppl}, Sec.~\ref{ED}). In sharp contrast to the significant Bragg peak at $\Gamma$ upon entering the FM phase for $J_{\pm}>0$ (blue symbols in Fig.~\ref{fig:fig4}(b)), no significant Bragg peak is observed at any momentum point for $J_{\pm}<0$ in either the spin structure factor $S(\mathbf{q})$ or the dimer structure factor $D(\mathbf{q})$. Although the system sizes accessible in ED are limited, the absence of any Bragg peak suggests that there is no conventional order, such as magnetic order or valence bond solid order, in this distinct phase.

To further probe the nature of this distinct phase, we compute the GMN for several representative subregions: the tripartite bowtie-212 (green points), and tripartite hexagonal truncations—hex-212 (yellow points). The GMN analysis is shown in the Fig.~\ref{fig:fig4}(d). In these three regimes, the GMN exhibits distinct behaviors. For $J_{\pm}>0.07076$, the system enters the FM phase, where finite GMN is detected in both loopy (bowtie-212) and non-loopy (hex-212) subregions. In the intermediate range, shown as the orange shaded region in Fig.~\ref{fig:fig4}, the BFG model realizes a $\mathbb{Z}_2$ QSL described at low energies by an effective four-spin exchange model. In this regime, no GMN is detected for any of the subregions considered. This absence of GMN within the $\mathbb{Z}_2$ QSL is consistent with a minimal multipartite entangled subregion larger than the partitions accessible in our analysis~\cite{Entanglement2025Song}. By contrast, in the green shaded region of Fig.~\ref{fig:fig4} (roughly $J_{\pm}<-0.1$), the GMN for bowtie-212 emerges near $J_{\pm} \sim -0.15$ and increases as $J_{\pm}$ decreases, indicating that the minimal multipartite entangled subregion shifts toward the bowtie geometry. 
Crucially, however, the GMN for the non-loopy hex-212 subregion remains zero throughout the entire $J_{\pm}<0$ regime, which suggests that the minimal multipartite entangled subregion in this regime is likely a loopy region, such as the bowtie-212. This observation is consistent with the conjecture that there is no non-loopy GMN in the regimes which hold gauge field descriptions, such as QSL phases~\cite{lyu2025multiparty,Entanglement2025Song}.

Taken together, the evidence from the QFI map, energy spectrum, structure factors, and GMN analysis all point to the existence of a distinct phase in the $J_{\pm}<0$ regime of the BFG model for $J_{\pm}<-0.10$. Going a step further, the absence of conventional order and the emergence of multipartite entanglement only in loopy subregions suggest that this distinct phase may be a different type of QSL from the $\mathbb{Z}_2$ QSL phase, which calls for further investigation.

\noindent\textcolor{blue}{\it Discussions.}--- Our results show that, depending on the type of the transition, i.e., beyond-Landau or Landau type, a clear signature in the QFI scaling behavior can be used to identify the QSL state and the associated fractionalization and emergent gauge field structure. For a conventional Landau-type (2+1)d XY transition, the exponent in Eq.~\eqref{eq:eq3} is $\Delta_Q > 0$. In contrast, for an unconventional (2+1)d XY$^\ast$ transition with enhanced scaling dimension due to fractionalization, the exponent becomes $\Delta_Q < 0$. For $\Delta_Q > 0$ (conventional case), the critical scaling of the QFI goes one way with system size $L$ or inverse temperature $\beta=1/T$. For $\Delta_Q < 0$ (unconventional case), the QFI goes oppositely with $L$ or $\beta$. This stark contrast in scaling direction bestows the dynamical scaling of the QFI with the power to distinguish novel QSLs and their unconventional QCPs from conventional phases. Our data confirm this signature, providing direct evidence of this distinction. 

Moreover, in the AFM regime of the BFG model, our ED results suggest the possible emergence of another QSL phase distinct from the $\mathbb{Z}_2$ QSL phase, as indicated by the QFI map, energy spectrum, structure factors, and GMN analysis, which would inspire further investigations. 

Finally, we emphasize that the QFI measures truly quantum correlations as the temperature-dependent form factor in Eq.~\eqref{eq:QFI} suppresses $T > \omega$ part of the structure factor. In other words, the QFI is significant only when temperature is smaller than a characteristic energy scale in the system, which corresponds to the quantum regime. 
This understanding and our demonstration of critical scaling in QFI near a beyond-Landau QCP provide useful guidance for future experimental studies of frustrated magnets such as kagome QSL candidate materials.

{\noindent\it Acknowledgement-----} 
We acknowledge inspiring discussions with William Witczak-Krempa, Shiliang Li and Cristian Batista on related topics. We acknowledge the support from the Research Grants Council (RGC) of Hong Kong (Project Nos. 17309822, C7037-22GF, 17302223, 17301924, 17301725), the ANR/RGC Joint Research Scheme sponsored by RGC of Hong Kong and French National Research Agency (Project No. A\_HKU703/22) and the State Key Laboratory of Optical Quantum Materials at HKU. We thank HPC2021 system under the Information Technology Services at the University of Hong Kong~\cite{hpc2021}, as well as the Beijing Paratera Tech Corp., Ltd~\cite{paratera} for providing HPC resources that have contributed to the research results reported within this paper. Z.Z. and Y.B.K. were supported by the Natural Sciences and Engineering Research Council of
Canada (NSERC) Grant No. RGPIN-2023-03296 and the Center of Quantum Materials at the University of Toronto. Computations at the University of Toronto were performed
on the Cedar and Fir clusters, which are hosted by the Digital Research Alliance of Canada. Z.Z. is further supported by the
Ontario Graduate Scholarship.

\bibliography{ref.bib}
\bibliographystyle{apsrev4-1}

\newpage
\clearpage
\begin{center}
	\textbf{\large Supplemental Material}
\end{center}
\setcounter{equation}{0}
\setcounter{figure}{0}
\setcounter{table}{0}
\setcounter{page}{1}
\setcounter{section}{0}

\makeatletter
\renewcommand{\theequation}{S\arabic{equation}}
\renewcommand{\thefigure}{S\arabic{figure}}
\setcounter{secnumdepth}{3}
\setcounter{page}{1}
\setcounter{equation}{0}
\setcounter{figure}{0}
\renewcommand{\theequation}{S\arabic{equation}}
\renewcommand{\thefigure}{S\arabic{figure}}
\setcounter{secnumdepth}{3}

\section{Quantum Monte Carlo}
\label{QMC}
\begin{figure}
	\centering
	\includegraphics[width=\columnwidth]{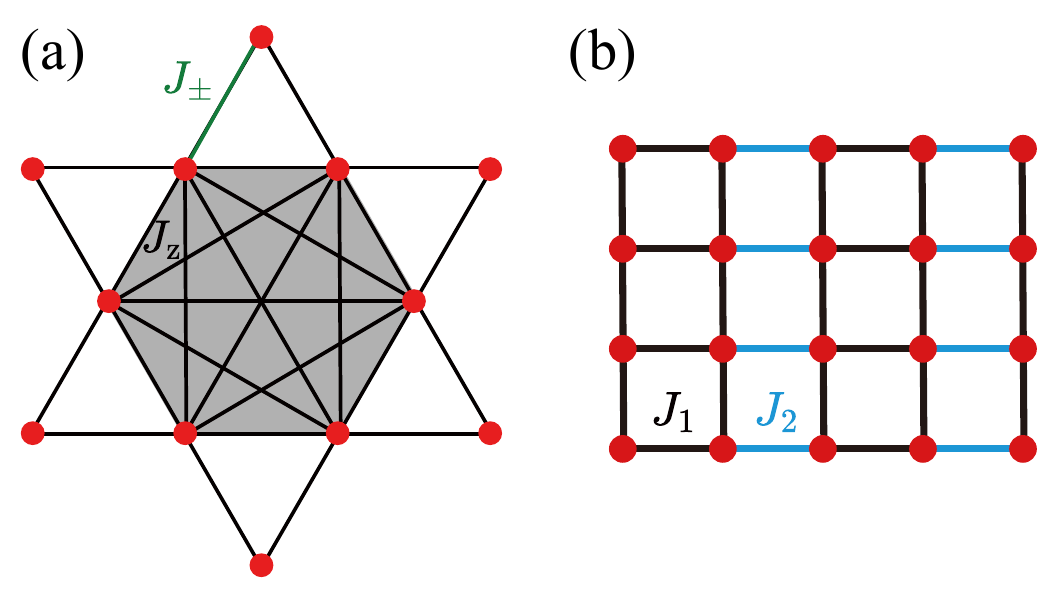}
	\caption{The schematic figure of the BFG model and the $J_1$-$J_2$ model. (a) The BFG model on the kagome lattice. The black bonds denote the Ising-type $J_z$ coupling, while the green bonds denote the $J_{\pm}$ coupling. (b) The easy-plane $J_1$-$J_2$ model on the square lattice. The black bonds denote the $J_1$ coupling and the blue bonds denote the $J_2$ coupling.}
	\label{fig:sm_lattice}
\end{figure}

In this section, we provide technical details of the quantum Monte Carlo (QMC) simulations for the Balents-Fisher-Girvin (BFG) model defined in Eq.~\ref{eq:Hamiltonian} of the main text as well as that of the easy-plane $J_1$-$J_2$ model defined in Eq.~\ref{eq:J1J2} of the main text. Their schematic figures of both models are shown in Fig.~\ref{fig:sm_lattice}. We simulate both of them using the stochastic series expansion (SSE) QMC method with directed loop updates~\cite{Computational2010Sandvik}.

Specifically, to handle the complexity of the BFG model and enter the $\mathbb{Z}_2$ quantum spin liquid (QSL) regime, we decompose the Hamiltonian into five-spin operators as in Refs.~\cite{wangTopological2017,Dynamical2018Sun,QSL2018Wang,Fractionalized2021wang} and and applied the balance condition but not the detailed balance condition solution for the directed loop update process~\cite{Markov2010Suwa}. Moreover, the multi-directed loop update method is also applied in the QSL regime ($J_{\pm}<0.7076$)\cite{zhouQuantum2025}. Meanwhile, the thermal annealing process is employed to further enhance the efficiency of the QMC simulations with the annealing step $\Delta \beta=1$ (where $\beta$ is the inverse temperature) and 10,000 Monte Carlo steps at each temperature. For the easy-plane $J_1$-$J_2$ model, we decompose the Hamiltonian into bond (two-spin) operators and apply the conventional directed loop update process~\cite{Computational2010Sandvik}.

\begin{figure}
	\centering
	\includegraphics[width=\columnwidth]{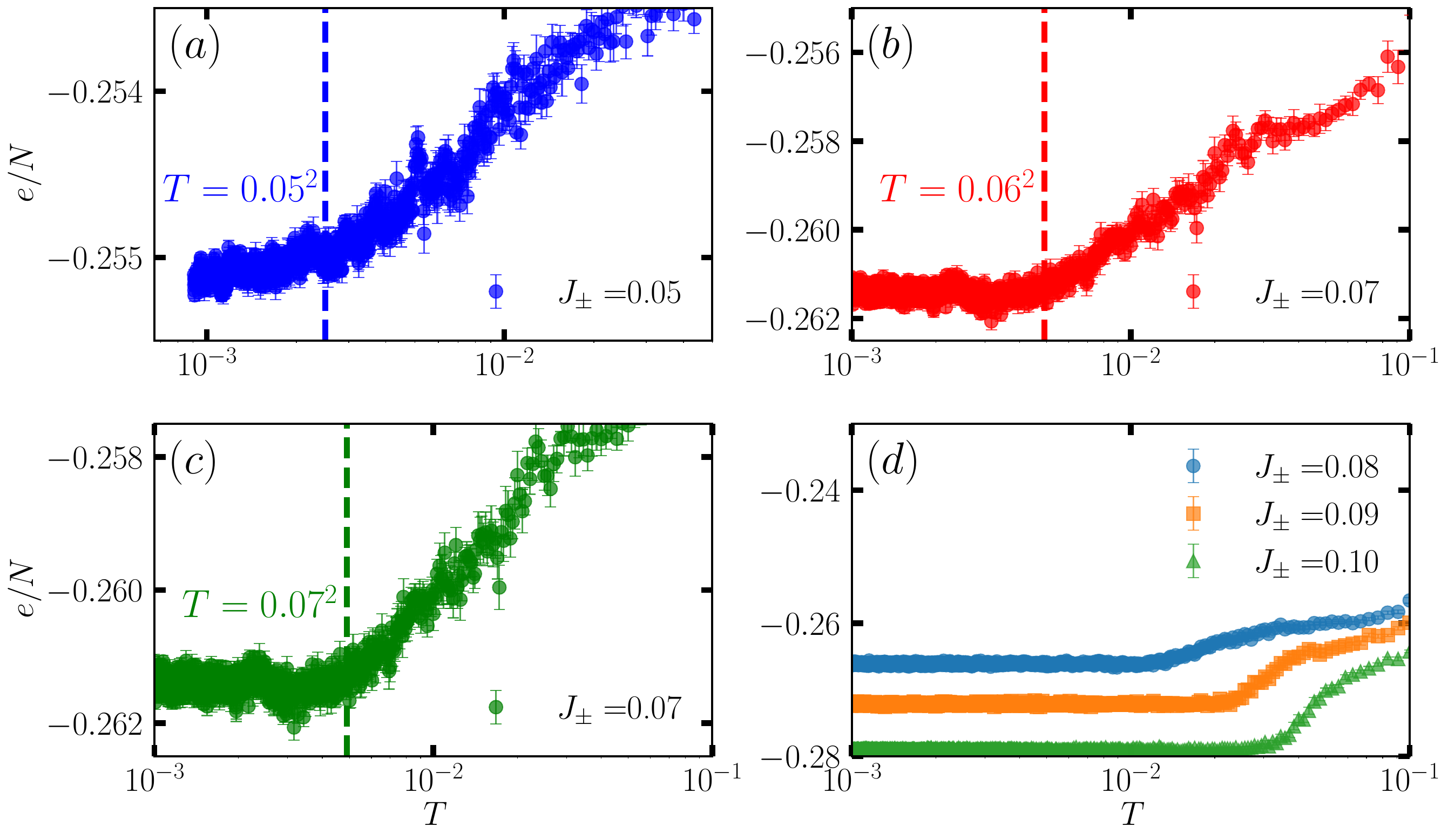}
	\caption{The energy per site $e/N$ obtained from QMC simulation for $L=6$. Panels (a), (b) and (c) correspond to $J_{\pm}=0.05$, $0.06$ and $0.07$, respectively. The dashed lines represent the temperature scale of entering the $\mathbb{Z}_2$ QSL state ($T \sim J_{\pm}^2$).  And panel (d) shows the $e/N$ as a function of temperature $T$ in the FM phase ($J_{\pm}>0.07076$).}
	\label{fig:enrgy_QMC}
\end{figure}

Figure~\ref{fig:enrgy_QMC} shows the energy per site $e/N$ obtained from QMC simulations for the BFG model with system size $L=6$ at different $J_{\pm}$. Panels (a), (b), and (c) correspond to $J_{\pm}=0.05$, $0.06$, and $0.07$, respectively. The dashed lines represent the temperature scale of entering the $\mathbb{Z}_2$ QSL state ($T \sim J_{\pm}^2$). $e/N$ decreases as the temperature decreases and enters a plateau below the temperature scale of $T \sim J_{\pm}^2$, indicating the system enters the $\mathbb{Z}_2$ QSL phase. Panel (d) shows $e/N$ as a function of temperature $T$ in the FM phase ($J_{\pm}>0.07076$). Here, $e/N$ also decreases as the temperature decreases, reflecting the FM order at low temperatures in a finite size.

During the QMC simulations, we also measure the imaginary-time spin-spin correlation functions $\langle S^\alpha_{\mathbf{q}}(\tau) S^\alpha_{-\mathbf{q}}(0) \rangle$ at various momentum points $\mathbf{q}$ and imaginary times $\tau$~\cite{Amplitude2021zhou,Dynamical2018Sun,Dynamical2018ma}. To access the real-frequency dynamics, we employ stochastic analytic continuation (SAC)~\cite{Progress2023shao,Stochastic1998sandvik,Nearly2017shao,Amplitude2021zhou} to extract the dynamical spin structure factor (DSSF) $A^\alpha(\mathbf{q}, \omega)$ from the imaginary-time data. The quantum Fisher information (QFI) density $f_Q$ is then computed using the obtained DSSF according to Eq.~\ref{eq:QFI}.

The QFI is employed to quantify the entanglement properties of a given state, as it provides a lower bound on multipartite entanglement, also known as entanglement depth~\cite{Fisher2012Hyllus,Tutorial2025Scheie,Witnessing2025Sabharwal}.
For a collective generator $\mathcal{O}=\sum_{i=1}^N \mathcal{O}_i$ with local eigenvalue range
$\Delta\lambda=\lambda_{\max}-\lambda_{\min}$, the entanglement depth can be detected by the following criterion:
\begin{equation}
\mathrm{nQFI}(\mathcal{O}):=\frac{f_Q(\mathcal{O})}{(\Delta\lambda)^2} > m\Longrightarrow
\text{entanglement depth } \ge m+1,\label{eq:nQFI}
\end{equation}
where $\mathrm{nQFI}$ denotes the normalized QFI. This criterion implies that the state is at least a $(m+1)$-producible state~\cite{hyllus2012fisher, toth2012multipartite, Witnessing2021Scheie, Tutorial2025Scheie, liu2016quantum}.

\begin{figure}
	\centering
	\includegraphics[width=\columnwidth]{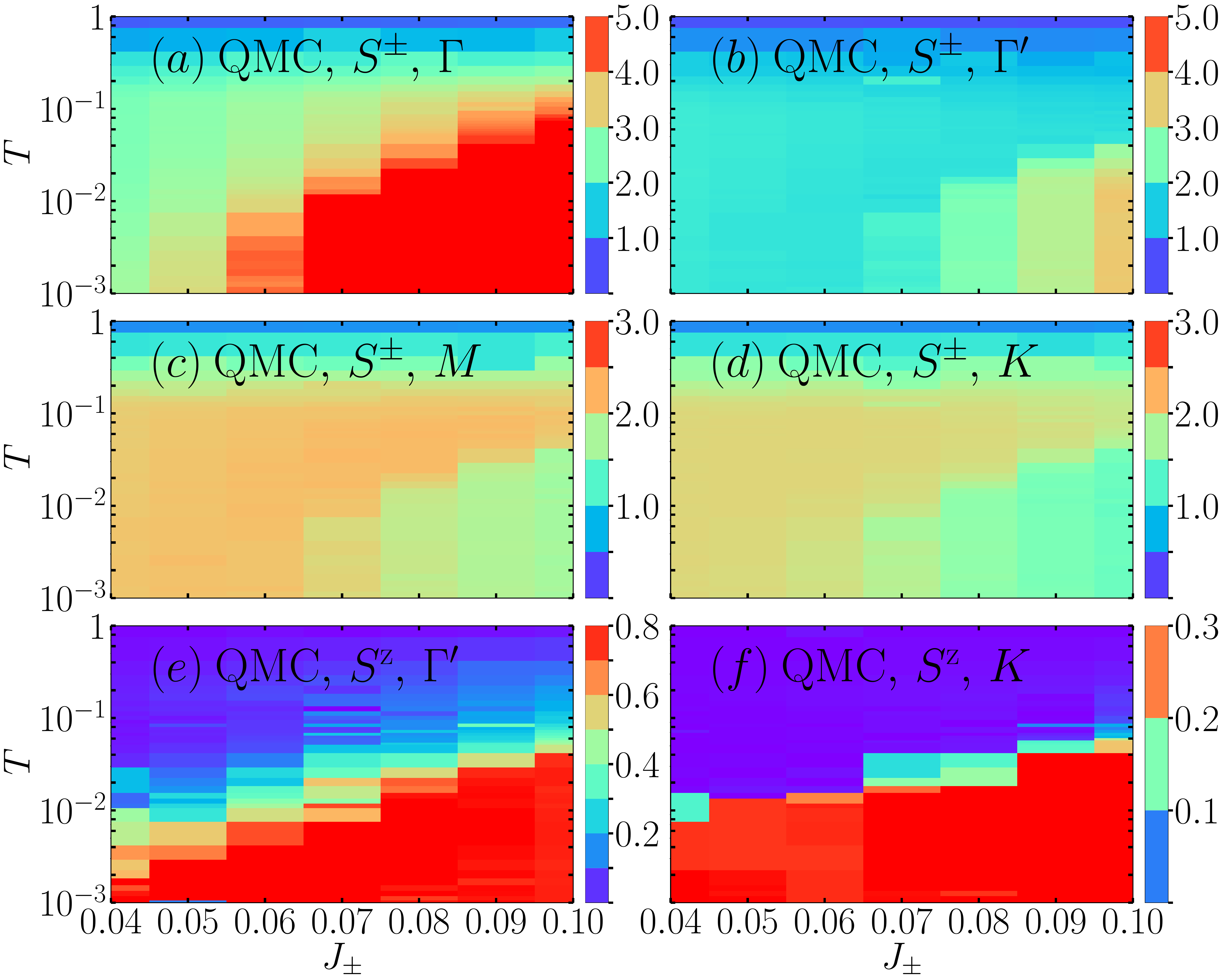}
	\caption{The QFI density $f_Q$ in the $S^{\pm}$ channel obtained from QMC as a function of $J_{\pm}$ and temperature $T$ with system size $L=6$ at different momentum points: (a) $\mathbf{q}=\Gamma$, (b) $\mathbf{q}=\Gamma^{\prime}$, (c) $\mathbf{q}=M$ and (d) $\mathbf{q}=K$. And panels (e-h) are the $S^{z}$ channel at the momentum points (e) $\mathbf{q}=\Gamma^{\prime}$ and (f) $\mathbf{q}=K$.}
	\label{fig:qfi_QMC}
\end{figure}

Figure~\ref{fig:qfi_QMC} presents the QFI density $f_Q$ obtained from QMC simulations for the BFG model with system size $L=6$ at different momentum points $\mathbf{q}$ as a function of $J_{\pm}$ and temperature $T$. Panels (a-d) correspond to the $S^{\pm}$ channel at momentum points $\Gamma$, $\Gamma^{\prime}$, $M$, and $K$, respectively. Panels (e-f) correspond to the $S^{z}$ channel at momentum points $\Gamma^{\prime}$ and $K$, respectively. In all panels, $f_Q$ increases as the temperature decreases. The QFI density $f_Q$ in the $S^{\pm}$ channel at the $\Gamma$ point (panel (a)) shows the highest value among all channels and momentum points, indicating the dominant role of $f_Q$ since QFI reflects the lower bound of multipartite entanglement in the system. Also, this channel serves as a good probe to observe the thermal phase diagram of the BFG model since the order vector is at $\Gamma$ point for the FM phase. Meanwhile, with respect to the ED results shown in Fig.~\ref{fig:qfi_Full} of the main text, we include the QMC results for the $S^{\pm}$ channel at momentum point $\Gamma^{\prime}$ and $S^{z}$ channel at momentum point $K$ in the main text to compare with the ED results. The other momentum points and channels are provided here for completeness.

\begin{figure*}
	\centering
	\includegraphics[width=\textwidth]{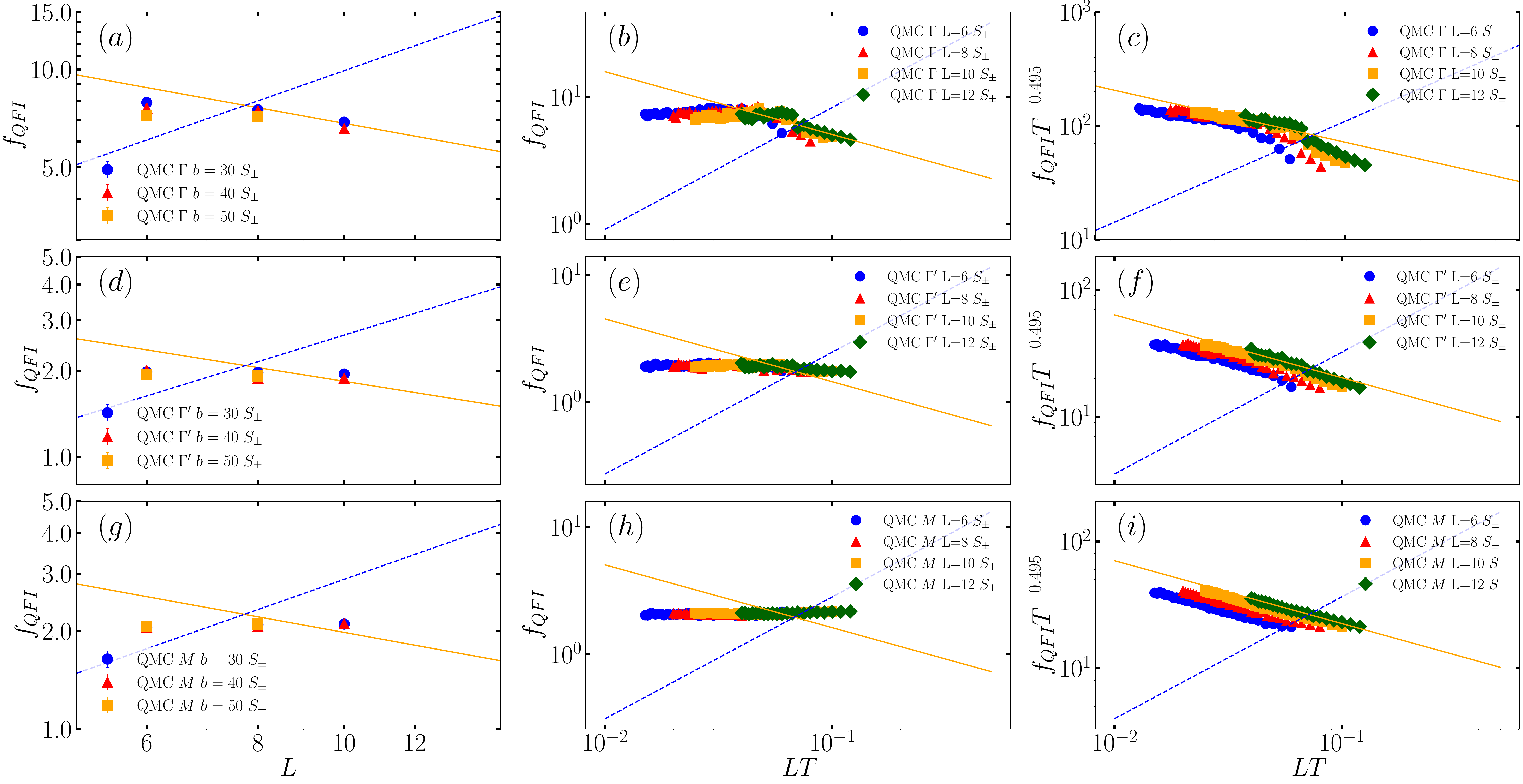}
	\caption{The scaling behavior of QFI density $f_Q$ is shown at momentum points $\Gamma$ (a-c), $\Gamma^{\prime}$ (d-f), and $M$ (g-i). For (a,d,g), $f_Q$ is plotted as a function of $L$ with fixed $T=1/(bL)$. For (b,e,h), $f_Q$ is plotted as a function of $LT$. For (c,f,i), $f_Q T^{-0.495}$ is plotted as a function of $LT$. The data are for system sizes ranging from $L=6$ to $L=12$, with $\beta\in[100,400]$. The orange solid lines correspond to a power-law with exponent $-0.495$ ((2+1)d XY$^\ast$ beyond-Landau QCP), while the blue dashed lines correspond to an exponent of $0.96$ ((2+1)d XY Landau QCP).}
	\label{fig:scaling_all}
\end{figure*}

To provide comprehensive results of the scaling behavior of the QFI, we also plot the finite size scaling following the formula that $f_Q = L^{\Delta_Q} \phi(1/b)$ with fixing $T=1/(bL)$ and $f_Q T^{\Delta_Q} = (TL)^{\Delta_Q} \phi(TL)$ at the critical point $J_{c,\pm}=0.07076$ and $\Delta_Q =-0.495$ in Fig.~\ref{fig:scaling_all}. Figure~\ref{fig:scaling_all} shows the scaling behavior of QFI density $f_Q$ at momentum points $\Gamma$ (a-c), $\Gamma^{\prime}$ (d-f), and $M$ (g-i). For (a,d,g), $f_Q$ is plotted as a function of $L$ with fixed $T=1/(bL)$ and $b$ ranging from $20$ to $50$. For (b,e,h), $f_Q$ is plotted as a function of $LT$. For (c,f,i), $f_Q T^{-0.495}$ is plotted as a function of $LT$. The data are for system sizes ranging from $L=6$ to $L=12$, with $\beta\in[100,400]$. The orange solid lines correspond to a power-law with exponent $-0.495$ ((2+1)d XY$^\ast$ beyond-Landau QCP), while the blue dashed lines correspond to an exponent of $0.96$ ((2+1)d XY Landau QCP). As shown in these figures, the QFI density $f_Q$ at $\Gamma$ point follows the expected scaling behavior with exponent $-0.495$, while the QFI density $f_Q$ at $\Gamma^{\prime}$ and $M$ points deviate from such scaling behavior, indicating that the dominant contribution to the QFI at the critical point arises from the $\Gamma$ point.

\begin{figure}
	\centering
	\includegraphics[width=\columnwidth]{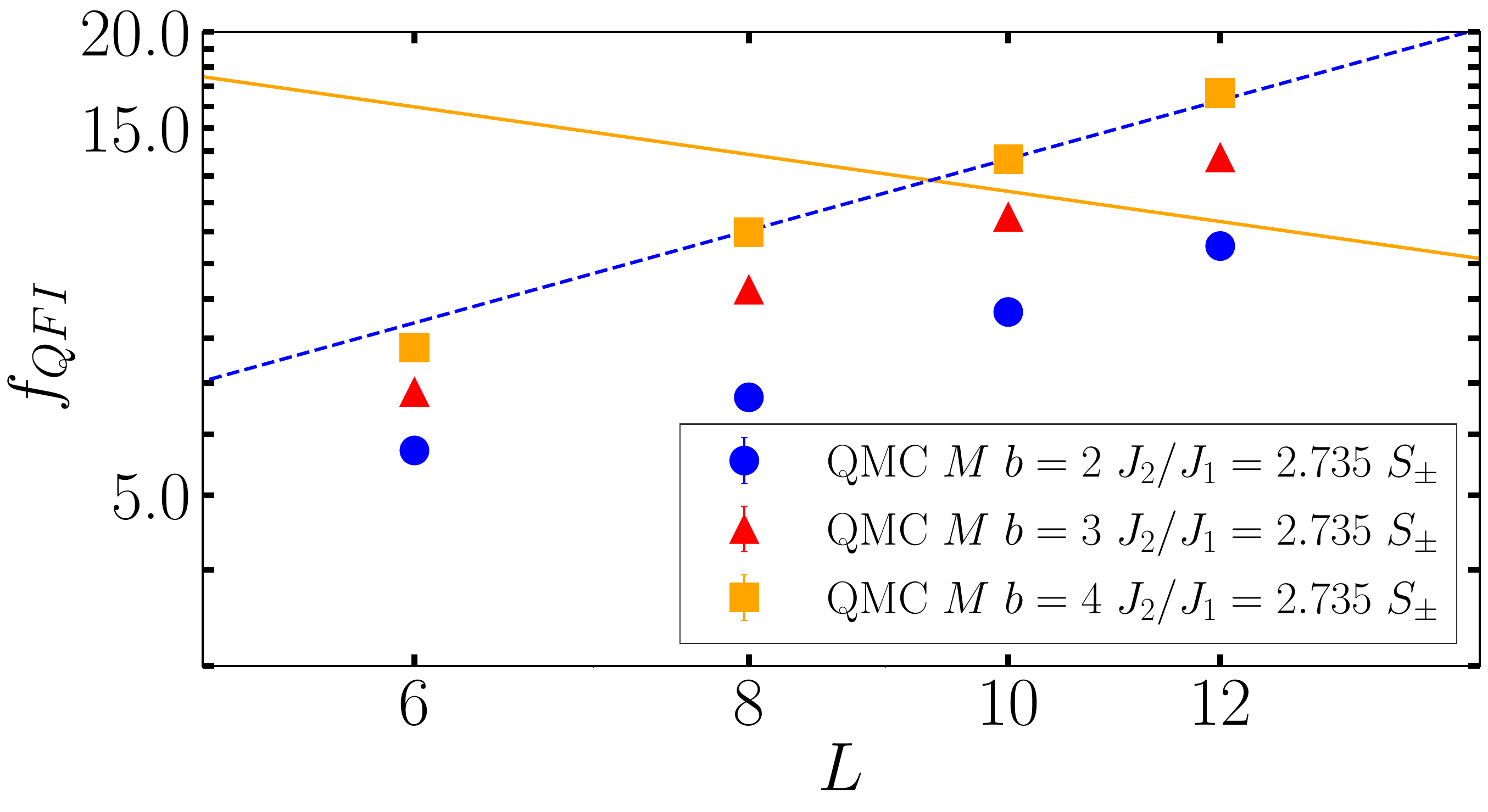}
	\caption{The QFI density $f_Q$ obtained from QMC on the $J_1$-$J_2$ model as a function of system size $L$ and temperature $T=1/(bL)$ at $\mathbf{q}=M$, with $L$ ranging from $6$ to $10$ and $b$ ranging from $2$ to $5$.}
	\label{fig:qfi_J1J2}
\end{figure}

Finally, in the Figure~\ref{fig:qfi_J1J2}, the scaling behavior of the $f_Q$ is plotted as a function of $L$ with fixed $T=1/(bL)$ and $b$ ranging from $2$ to $4$ for $S^{\pm}$ channel at $M=(\pi,\pi)$ points. Here, system size $L$ ranges from $6$ to $12$ with the same as the BFG model. As $b$ increases, the temperatures for these system sizes also decrease and the values of $f_Q$ increase. The data follow a power-law behavior with an estimated exponent of $0.96$, and the quality of the fit improves for larger $b$. Therefore, in the main text, we choose $b=4$ to show the scaling behavior of $f_Q$ for the easy-plane $J_1$-$J_2$ model at the XY critical point.

\section{exact diagonalization}
\label{ED}

Our finite-temperature exact diagonalization calculations are based on the microcanonical
thermal pure quantum (mTPQ) formulation~\cite{thermal2012sugiura,canonical2013sugiura, iitaka2004randomphase, long2003mclm, hams2000eigdensity}.
We begin from an $T\!\to\!\infty$ random vector in the working Hilbert space,
\begin{equation}
    |\psi_0\rangle=\sum_i c_i |i\rangle,
\end{equation}
where $\{c_i\}$ are independent random complex coefficients.  Given a Hamiltonian $H$,
the $k$th mTPQ state is obtained by repeatedly applying an energy-shifted Hamiltonian and
renormalizing at each step,
\begin{equation}
    |\psi_k\rangle
    = \frac{(L-H)\,|\psi_{k-1}\rangle}{\|(L-H)\,|\psi_{k-1}\rangle\|},
    \qquad L > E_{\max}(H),
    \label{eq:mtpq_iter}
\end{equation}
where $E_{\max}(H)$ is the largest eigenvalue of $H$.  The internal energy at iteration $k$
is estimated as
\begin{equation}
    E_k=\frac{\langle \psi_k |H|\psi_k\rangle}{\langle \psi_k |\psi_k\rangle},
\end{equation}
and the corresponding inverse temperature is approximated by~\cite{thermal2012sugiura,canonical2013sugiura}
\begin{equation}
    \beta_k \simeq \frac{2k}{L-E_k},
    \label{eq:beta_mtpq}
\end{equation}
noting that $k$ scales extensively with system size for fixed $\beta$.  In the thermodynamic
limit, a single realization of $|\psi_k\rangle$ is typical and reproduces equilibrium expectation
values, so static observables are evaluated as
\begin{equation}
    \langle O\rangle_{\beta_k} \approx
    \frac{\langle \psi_k|O|\psi_k\rangle}{\langle \psi_k|\psi_k\rangle}.
\end{equation}
We therefore leverage this formalism to extract the specific heat via evaluating $\langle H^2\rangle_{\beta_k}-E_k^2$.
For finite clusters, we average over 8 independent random seeds to
reduce residual statistical fluctuations. 

We evaluate finite-temperature spectra from the mTPQ state $|\psi_k\rangle$ via the broadened
resolvent associated with an operator $\mathcal{O}$~\cite{jaklic1994ftlm, prelovsek2013lanczoschapter},
\begin{equation}
    G_{\mathcal{O}}(z;\beta_k)
    =
    \frac{\langle \psi_k|\mathcal{O}^\dagger\,(z-\widetilde{H})^{-1}\mathcal{O}|\psi_k\rangle}
         {\langle \psi_k|\psi_k\rangle},
    \label{eq:resolvent_def}
\end{equation}
where $
    \widetilde{H}\equiv H-E$, $
    z=\omega+i\eta$. $\eta>0$ sets the Lorentzian broadening, and $E$ is taken to be the energy of the ground state, which gives the correct energy offset for the spectrum. In practice, the ground state energy is obtained via a separate Lanczos process.
The corresponding spectral density is
\begin{equation}
    I_{\mathcal{O}}(\omega;\beta_k) = -\frac{1}{\pi}\,\mathrm{Im}\,G_{\mathcal{O}}(\omega+i\eta;\beta_k).
    \label{eq:spectral_def}
\end{equation}

Define the Lanczos seed and its prefactor
\begin{equation}
    |f_0\rangle = \frac{\mathcal{O}|\psi_k\rangle}{\|\mathcal{O}|\psi_k\rangle\|},
    \qquad
    \mu_0 \equiv \frac{\|\mathcal{O}|\psi_k\rangle\|^2}{\langle \psi_k|\psi_k\rangle}.
    \label{eq:seed_def}
\end{equation}
Applying the Lanczos tridiagonalization to $\widetilde{H}$ with starting vector $|f_0\rangle$
yields a tridiagonal representation characterized by diagonal coefficients $\{a_n\}$ and
off-diagonal coefficients $\{b_{n+1}\}$~\cite{gagliano1987,hallberg1995,prelovsek2013lanczoschapter}.

In this basis, the Krylov-space approximation to the resolvent matrix element takes the standard
continued-fraction form~\cite{gagliano1987,hallberg1995}
\begin{equation}
    G_{\mathcal{O}}(z;\beta_k)
    \approx
    \mu_0\,
    \frac{1}{z-a_0-\cfrac{b_1^2}{z-a_1-\cfrac{b_2^2}{z-a_2-\ddots}}},
    \label{eq:continued_fraction}
\end{equation}
which is truncated after a finite number of Lanczos steps $n_{\mathrm{L}}$.
Equations~\eqref{eq:spectral_def} and \eqref{eq:continued_fraction} then give $I_{\mathcal{O}}(\omega;\beta_k)$ directly. 

We implement the above mTPQ--Lanczos workflow with GPU acceleration on a 27-site ($3\times 3$) kagome torus with periodic boundary conditions. Among $N_s=27$ kagome tori, this geometry is commonly used as a high-symmetry cluster with comparatively large shortest winding loops (torus diameters), which helps mitigate---though does not eliminate---finite-size artifacts. Such artifacts are known to be particularly pronounced on the Kagome lattice, where gaps and low-energy spectra can vary substantially across cluster shapes and sizes \cite{Lauchli2011Ground,Singh2008Triplet,Nakano2011Numerical}. 

In addition, typicality-based finite-temperature estimators rely on stochastic sampling over random initial vectors; at finite size, this induces a seed-to-seed spread that is expected to be largest near thermal crossover regimes \cite{Jaklic1994Lanczos,Prelovsek2013Ground,Misawa2020Asymmetric, zhouQuantum2025}. In our microcanonical implementation, finite-$T$ dynamical response functions are evaluated from a single appropriately chosen energy-window state; this approach becomes exact in the thermodynamic limit and yields increasingly smooth spectra with increasing system size, while smaller clusters can retain residual statistical fluctuations \cite{long2003mclm,Okamoto2018Accuracy}. Consistent with these expectations, we find $S(\mathbf{q},\omega;T)$ (and the QFI derived from it) to be stable over most temperatures, with the largest variance confined to an intermediate crossover window. We therefore report data averaged over 16 independent mTPQ realizations. 


\begin{figure}
	\centering
	\includegraphics[width=\columnwidth]{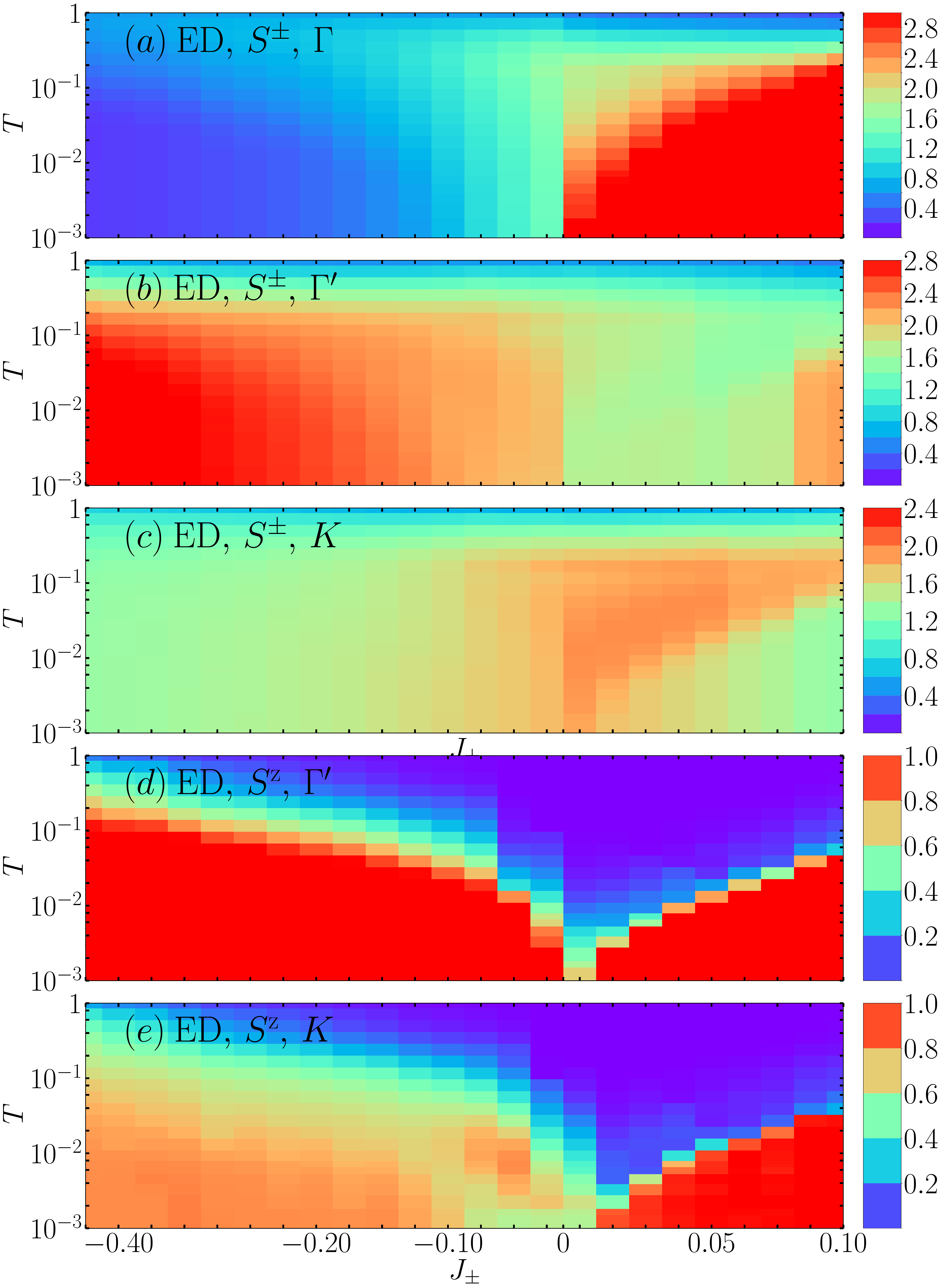}
	\caption{The QFI density $f_Q$ obtained from ED on the BFG model as a function of both temperature $T$ and $J_{\pm}$ for system size $L_x\times L_y=3\times 3$ at momentum points (a) $S^{\pm}$ at $\Gamma$, (b) $S^{\pm}$ at $\Gamma^{\prime}$, and (c) $S^{\pm}$ at $K$. Meanwhile, panels (d) and (e) show the QFI density $f_Q$ in the $S^z$ channel at momentum points $\Gamma^{\prime}$ and $K$, respectively.}
	\label{fig:qfi_ED_3x3}
\end{figure} 

The energy and specific heat per site, \(C_v\), obtained from exact diagonalization on the \(L=3\times3\) cluster are shown in Fig.~\ref{fig:eng_all} for representative values of \(J_{\pm}\), with the insets highlighting the lowest-temperature \(C_v\) peak. The different panels correspond to distinct regimes of the phase diagram: (a) a putative distinct QSL, (b) the crossover out of the \(\mathbb{Z}_2\) QSL, (c,e) the \(\mathbb{Z}_2\) QSL, (d) the Ising limit (\(J_{\pm}=0\)), and (f) the FM phase. In contrast to the FM phase, whose specific heat exhibits a pronounced peak associated with the onset of FM order, the \(\mathbb{Z}_2\) QSL regime in panels (c,e) displays a characteristic two-peak structure, reflecting crossover scales associated with spinon and vison excitations. At lower \(J_{\pm}\), however, this two-peak structure becomes progressively suppressed and eventually evolves into a three-peak profile, suggesting the emergence of a phase distinct from the conventional \(\mathbb{Z}_2\) QSL. This interpretation is consistent with our complementary analyses based on the QFI and GMN.

\begin{figure}
	\centering
	\includegraphics[width=\columnwidth]{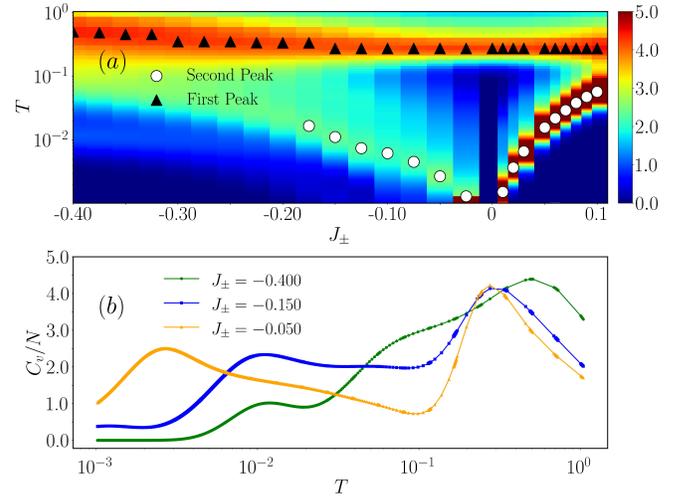}
  	\caption{ED results of the specific heat density $C_v/N$ on the AFM kagome model. (a) presents $C_v/N$ as a function of both $J_{\pm}$ and temperature $T$, while (b) shows line cuts of (a) at $J_{\pm}=-0.05$ for the $\mathbb{Z}_2$ QSL phase regime, at $J_{\pm}=-0.15$ in the crossover region, and at $J_{\pm}=-0.40$ for the green regime. The low-temperature (second) peak in (a) is highlighted by the white circle point while the high-temperature (first) peak is marked by the black triangle point.}
	\label{fig:Sfig_cv}
\end{figure}

Finally, the low-energy spectrum in Fig.~\ref{fig:fig4} is computed using the Krylov Schur algorithm per total $S^z$ sector per momentum symmetry sector.

\begin{figure}
    \centering
\includegraphics[width=\linewidth]{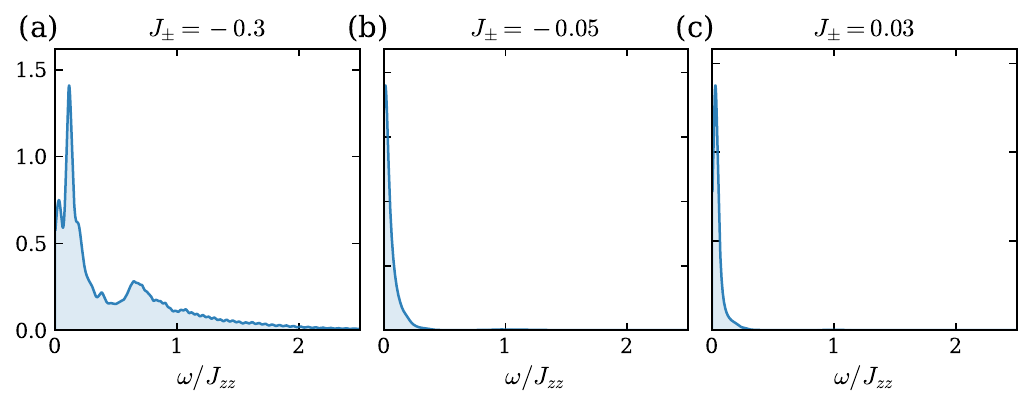}
    \caption{Dynamical spin structure factor $S^{zz}(\mathbf{q}=K,\omega)$ at $T=0.001$ for $J_\pm=-0.30,-0.05,0.03$ (a,b,c)}
    \label{fig:Supp_Sz_spectrum}
\end{figure}

\begin{figure*}
	\centering
	\includegraphics[width=\textwidth]{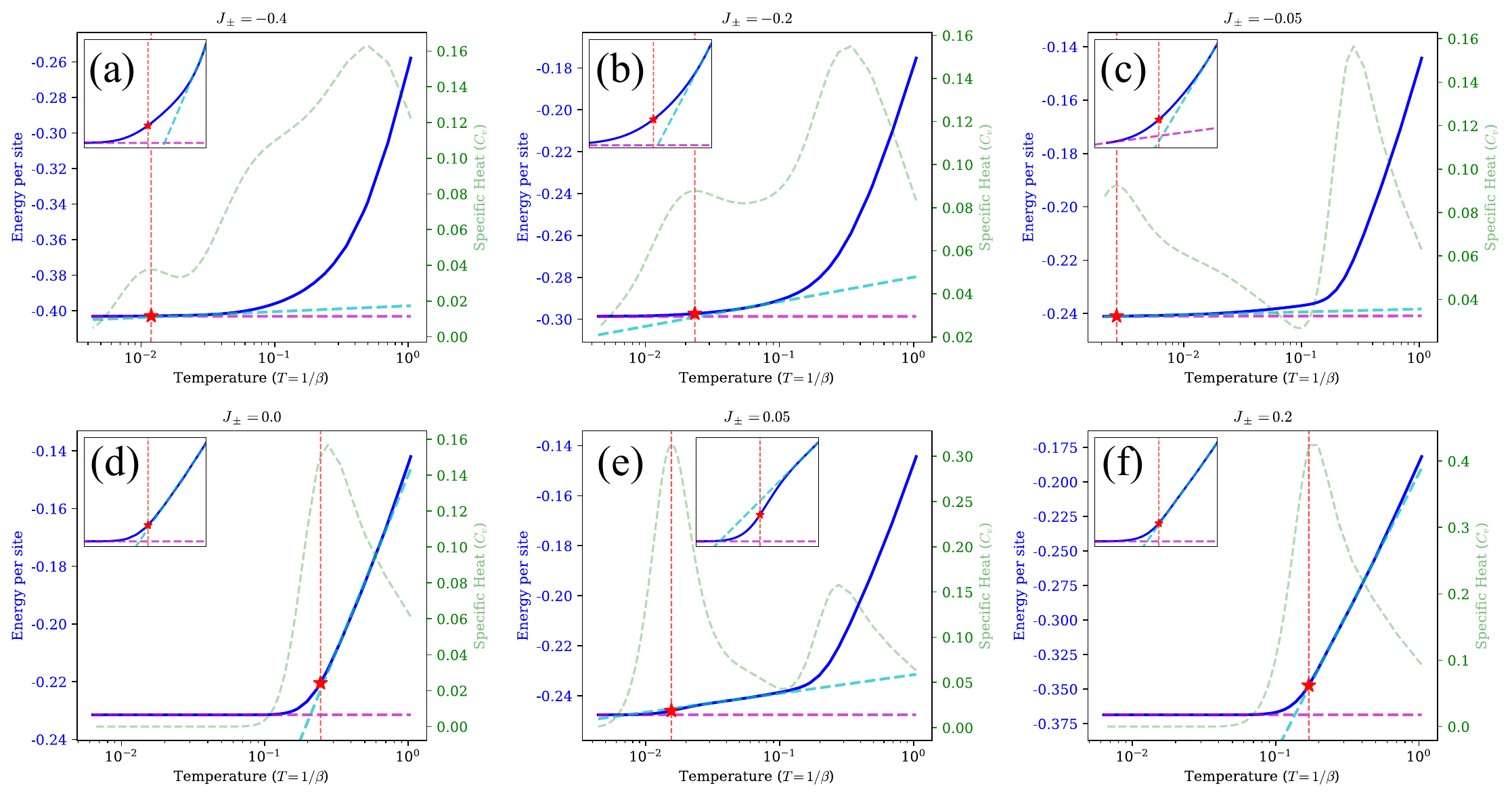}
	\caption{The energy per site and the specific heat $C_v$ for different $J_{\pm}$. Panels (a)-(d) correspond to $J_{\pm}=-0.4,-0.2,-0.05$, and $0.0$, respectively, while (e) and (f) correspond to $J_{\pm}=0.05$ and $0.2$, respectively. The insets of each panel show the zoomed energy behavior corresponding to the lowest specific heat peak. Panel (a) is in the distinct QSL regime while (b) is in the crossover from the $\mathbb{Z}_2$ QSL phase to the distinct QSL regime. Panels (c) and (e) are in the $\mathbb{Z}_2$ QSL phase while (f) is in the FM phase. Panel (d) refers to the Ising limit ($J_{\pm}=0$).}
	\label{fig:eng_all}
\end{figure*}

In addition to Fig.~\ref{fig:fig4} in the main text, we also present the dynamical spin structure factor $S^{zz}(\mathbf{q}=K,\omega,T=0.001)$ for $J_\pm=-0.30$, $-0.05$, and $0.03$, corresponding respectively to the distinct spin-liquid regime and the $\mathbb{Z}2$ spin-liquid regime, as shown in Fig.~\ref{fig:Supp_Sz_spectrum}. In the $\mathbb{Z}2$ spin-liquid regime, the $\langle S^z S^z \rangle$ channel is expected to probe gapped vison excitations. Consistent with this expectation, for $J\pm=-0.05$ and $J\pm=0.03$ the spectrum exhibits a pronounced low-frequency peak. By contrast, deep in the $J_\pm=-0.30$ regime, this clear vison-like signature disappears, and the spectrum instead becomes broad, with substantial weight extending to higher frequencies. This qualitative change is consistent with the picture that the emergent gauge structure in this phase differs from the $\mathbb{Z}_2$ bowtie-vison structure supported in the BFG model.

\section{Genuine Multipartite Negativity}
\label{GMN}
A foundational concept in multipartite quantum systems is the distinction between separable and entangled states. Consider a pure quantum state $|\psi\rangle$ of $n$ parties living in the composite Hilbert space $\mathcal{H} = \mathcal{H}_1 \otimes \cdots \otimes \mathcal{H}_n$. The state is fully separable if it can be expressed as a simple tensor product of the states of its individual subsystems: $|\psi\rangle = |\psi_1\rangle \otimes |\psi_2\rangle \otimes \cdots \otimes |\psi_n\rangle$. Any pure state that cannot be written in this factorized form is, by definition, entangled.
Moving beyond simple entanglement, a more complex structure arises in systems with three or more parties: genuine multipartite entanglement (GME). A state possesses GME if it is not biseparable. A state is considered biseparable if it can be expressed as a probabilistic mixture of states, where each state in the mixture is separable across at least one possible bipartition of the system. For instance, in a three-party system involving subsystems $A$, $B$, and $C$, a biseparable state $\rho^{\text{bsep}}$ takes the general form:
$ \rho^{\text{bsep}} = p_1 \rho_{A|BC}^{\text{sep}} + p_2 \rho_{B|AC}^{\text{sep}} + p_3 \rho_{C|AB}^{\text{sep}} $.
Here, each $\rho^{\text{sep}}$ is a state that is separable with respect to a specific cut (e.g., $A$ separated from the combined $BC$ system), and $p_i$ are probabilities summing to one. A state that cannot be represented in this convex-sum form is genuinely multipartite entangled.

To quantify GME, one can use computable entanglement monotones. One such measure is the Genuine Multipartite Negativity (GMN)~\cite{Guhne2011Taming,Hofmann2014Analytical}. The construction of GMN begins with the concept of negativity for a single bipartition. For any given division of the total system into two parts, $M$ and its complement $\overline{M}$, the negativity is defined as ${N}_{M|\overline{M}}(\rho) = \frac{1}{2} ( |\rho^{T_M}|_1 - 1 )$. This value is calculated from the partial transpose of the density matrix $\rho$ with respect to the subsystem $M$, where $|\cdot|_1$ denotes the trace norm. A positive negativity indicates entanglement across that specific partition.
From this, we define the minimum negativity, ${N}_{\text{min}}(\rho)$, as the smallest negativity found across all possible bipartitions of the system:
${N}_{\text{min}}(\rho) = \min_{M|\overline{M}} {N}_{M|\overline{M}}(\rho)$.
This quantity essentially captures the ``weakest link" of entanglement in the state.
However, for mixed states, this minimum negativity is not by itself a perfect entanglement monotone. To address this, the GMN, denoted as $\mathcal{N}(\rho)$, is formally defined through a convex-roof extension of the minimum negativity:
\begin{equation}
\mathcal{N}(\rho)=\min_{{p_k,\rho_k}} \sum_k p_k {N}_{\text{min}}\left(\rho_k\right),
\label{eq:GMN_definition}
\end{equation}
where the minimization is performed over all possible decompositions of the state $\rho$ into an ensemble of pure states ${p_k, \rho_k}$ such that $\rho=\sum_k p_k \rho_k$. This optimization ensures that the measure properly accounts for the convex structure of the set of separable states.
A key feature of GMN is that a positive value, $\mathcal{N}(\rho) > 0$, serves as a sufficient condition for the presence of genuine multipartite entanglement. It is important to note, however, that this condition is not necessary; some GME states may have a GMN of zero. For the simpler case of a two-party system, the GMN simplifies and becomes equivalent to the standard negativity measure.

A significant advantage of Genuine Multipartite Negativity (GMN) is that its value, as defined in Eq.~(\ref{eq:GMN_definition}), can be determined efficiently by solving a specific type of optimization problem known as a semidefinite program (SDP). This computational method circumvents the challenging minimization over all pure state decompositions. The SDP formulation for the GMN of a given state $\rho$ is expressed as follows:
\begin{equation}
\begin{aligned}
\mathcal{N}(\rho)&=-\min \operatorname{tr} (\rho W) \\
\text { subject to } & W=P_m+Q_m^{T_m} \\
& 0 \leqslant P_m  \\
& 0 \leqslant Q_m \leqslant I \text {  for all bipartitions } m \mid \bar{m}
\end{aligned}    
\end{equation}
In this formulation, the minimization is performed over the operator $W$, which acts as an entanglement witness. The constraints dictate that this witness must be decomposable across every possible bipartition $m|\bar{m}$ of the system. Specifically, for each bipartition, $W$ must be expressible as the sum of a positive semidefinite operator $P_m$ and the partial transpose of another operator $Q_m$. The operator $Q_m$ is itself constrained to be positive semidefinite and bounded by the identity operator $I$. The notation $A \geqslant B$ means that the operator $A-B$ is positive semidefinite, and $T_m$ denotes the partial transposition with respect to one of the subsystems in the bipartition $m|\bar{m}$.

For the practical computation of the GMN, our methodology is adapted from the MATLAB package pptmixer, originally introduced in Ref.~\cite{Guhne2011Taming}. The original constraint $P_m \leqslant I$ is removed as suggested in Ref.~\cite{Hofmann2014Analytical} for a more direct and physically meaningful interpretation of the measure.

We adopt the YALMIP toolbox~\cite{YALMIP2004} as a high-level modeling interface. For the numerical solution of these optimization problems, we employ the high-performance MOSEK solver~\cite{mosek}.

\begin{figure}
	\centering
	\includegraphics[width=\columnwidth]{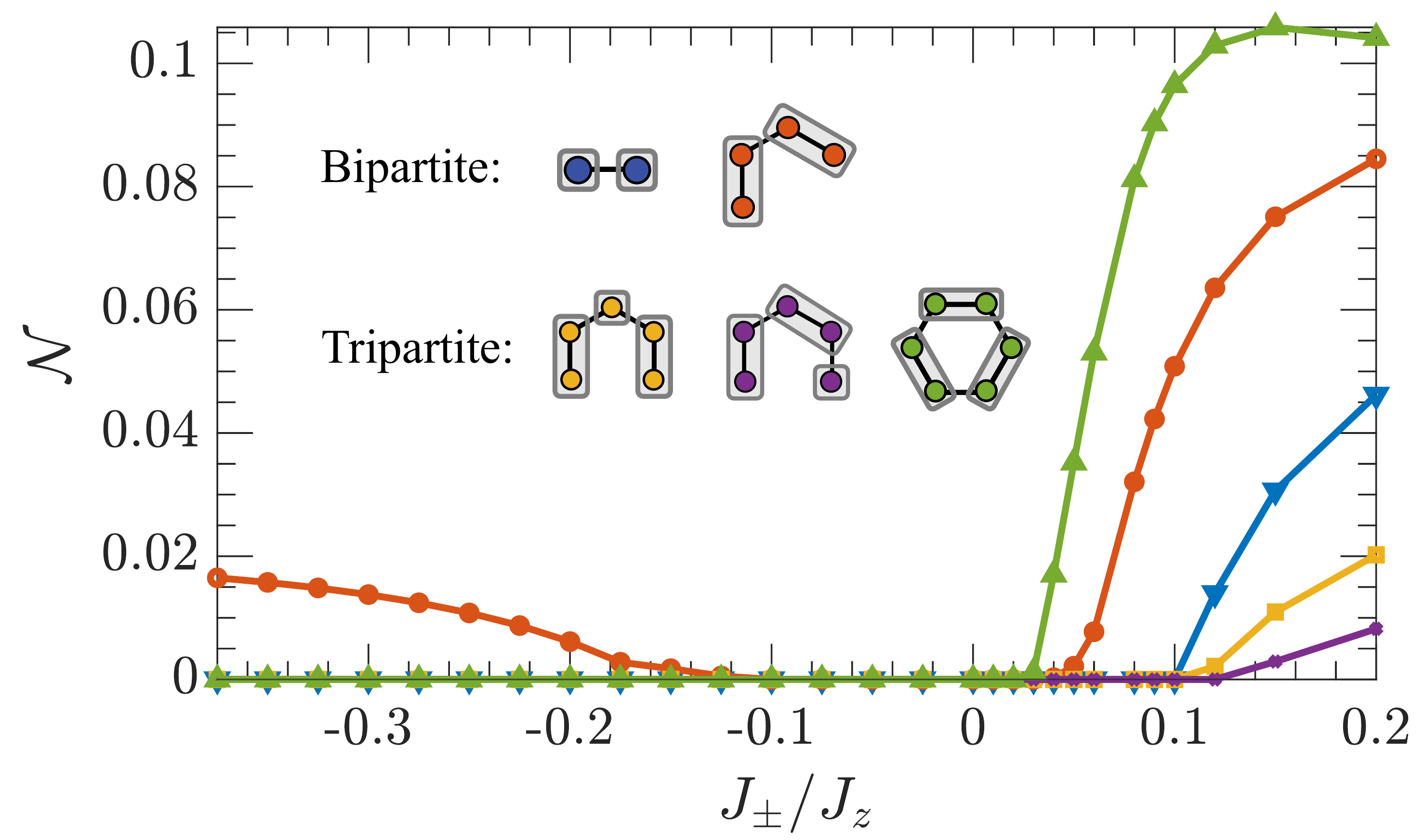}
	\caption{The GMN for the BFG model evaluated across different partition schemes on an $L=3\times3$ cluster. The inset illustrates the various partitions, with the color of each scheme matching its corresponding curve in the main plot.}
	\label{fig:sm_GMN}
\end{figure} 

Figure~\ref{fig:sm_GMN} presents the GMN for the BFG model under additional partition schemes (beyond those in the main text) on an $L = 3 \times 3$ cluster. In addition to this multipartite entanglement measure, we also show the bipartite negativity for two and four adjacent spins. The bipartite negativity for four adjacent spins forming a hexagon remains zero near $J_\pm = 0$ within the $\mathrm{Z}_2$ QSL phase. A finite value emerges around $J_\pm \approx -0.1$ as $J_\pm$ becomes more negative. This onset coincides with that of the tripartite bowtie GMN presented in the main text, providing further evidence for a phase transition.

For the tripartite GMN shown in Fig.~\ref{fig:sm_GMN}, we do not detect any finite signal in the QSL phases at negative $J_\pm$, including the “loopy hexagon” partition, although such contributions do appear in the ferromagnetic phase.

\newpage

\end{document}